\documentclass[a4paper,11pt]{article}
\usepackage[toc,page]{appendix}
\usepackage{jheppub}
\usepackage{setspace}
\usepackage{amssymb}
\usepackage{amsmath,amsfonts}
\usepackage{graphicx}
\usepackage{mathrsfs}
\usepackage[vcentermath]{youngtab}
\usepackage{color}
\usepackage{tikz}
\usetikzlibrary{shapes,arrows,shadows}
\numberwithin{equation}{section}
\usepackage{bbm}

\def\bea{\begin{eqnarray}}
\def\eea{\end{eqnarray}}

\newcommand{\eq}{\begin{equation}}
\newcommand{\en}{\end{equation}}
\newcommand{\eqn}{\begin{eqnarray}}
\newcommand{\enn}{\end{eqnarray}}
\newcommand{\nn}{\nonumber }
\newcommand{\beq}{\begin{equation}}
\newcommand{\eeq}{\end{equation}}

\usepackage[normalem]{ulem}

\begin{document}
\begin{center}
{\Large \bf Massless Conformal Fields in Ten Dimensions , Minimal Unitary Representation of $E_{7(-25)}$  and  Exceptional Supergravity} \\
\vspace{2cm}
{\large \bf Murat G\"unaydin}

{\it
Institute for Gravitation and the Cosmos,\\
Physics Department, \\
Pennsylvania State University \\
University Park, PA 16802, USA} \\
\vspace{0.5cm}
mgunaydin@psu.edu \\
\vspace{1cm}
{\bf Abstract} \\
\end{center} 
Minimal unitary representation of $SO(d,2)$ and its deformations describe all the conformally massless fields in $d$ dimensional Minkowskian spacetimes.  In critical dimensions these spacetimes admit extensions with twistorial coordinates plus a dilatonic coordinate to causal spacetimes coordinatized by Jordan algebras $J_3^{\mathbb{A}}$ of degree three over the four division algebras $\mathbb{A}=\mathbb{R} , \mathbb{C} , \mathbb{H} , \mathbb{O} $. We study the minimal unitary representation (minrep) of the conformal group $E_{7(-25)}$ of the  spacetime coordinatized by the exceptional Jordan algebra $J_3^{\mathbb{O}}$ defined over  octonions $\mathbb{O}$. 
We show that the minrep  of $E_{7(-25)}$  decomposes into infinitely many massless representations of the ten dimensional conformal group $SO(10,2)$.  Corresponding  conformal fields transform as symmetric tensors  in spinor indices of $SO(9,1)$ subject to certain constraints. Even and odd tensorial fields describe bosonic and fermionic conformal fields , respectively. Each irrep of $SO(10,2)$ comes with infinite multiplicity and fall into a unitary representation  of an $SU(1,1)$ subgroup that commutes with  $SO(10,2)$. 
The noncompact generators transforming in spinor representation $16 $ of $SO(10)$  interpolate between  the bosonic and fermionic representations  and hence act like "bosonic supersymmetry" generators.  These results truncate to similar results in the lower critical space-time dimensions. We also  give the decomposition of the minrep of  $E_{7(-25)}$ with respect to the subgroup $SO^*(12)\times SU(2)$ with $SO^*(12) $ acting as the conformal  group of the spacetime coordinatized by the $J_3^{\mathbb{H}}$ defined over quaternions $\mathbb{H}$.   Group $E_{7(-25)}$ is also the U-duality group of the exceptional $N=2$ Maxwell-Einstein supergravity in four dimensions. We discuss the relevance  of our results to the composite scenario that was proposed for  the exceptional supergravity so as to accommodate the families of quarks and leptons of the standard model as well as  to the proposal that $E_{7(-25)}$  acts as spectrum generating symmetry group of  the $5d$ exceptional supergravity. \\

{\it Contribution to Stanley Deser memorial volume " Gravity, Strings and Beyond" }
\newpage 
\tableofcontents

\newpage
\section{Introduction}
Quasiconformal realizations of noncompact groups was  discovered in \cite{Gunaydin:2000xr} and developed further in \cite{Gunaydin:2007qq,Gunaydin:2009dq,Gunaydin:2009zza,Gunaydin:2005gd,Gunaydin:2005zz}. 
By quantizing its geometric quasiconformal realization the minimal unitary representation (minrep) of the split exceptional group $E_{8(8)}$ was constructed in \cite{Gunaydin:2001bt}. Later minimal unitary representation of $E_{8(-24)}$ with the maximal subgroup $E_7\times SU(2)$  was constructed using the quasiconformal approach and truncated to the minreps of its subgroups, in particular $E_{7(-25)}$ and $E_{6(-14)}$ in \cite{Gunaydin:2004md}. 

The minimal unitary representation of a noncompact group is defined as the representation over an Hilbert space of functions of smallest number of variables possible. They were introduced by Joseph\cite{Joseph:1974} who was motivated by the work of physicists on spectrum generating algebras and were studied by mathematicians over the years\cite{MR644845,MR1159103,MR1372999,MR1278630,MR1267034,MR1889991,MR1737731}.

   A unified approach  to the minimal unitary representations of noncompact Lie groups and Lie superalgebras was formulated  in \cite{Gunaydin:2006vz} by quantizing their quasiconformal realizations. Using this method it was shown that minrep of $SO(d,2)$ and its deformations describe all the massless conformal fields in $d$ dimensional Minkowskian spacetimes\cite{Fernando:2015tiu}. The minrep describes a conformal scalar in all dimensions. One finds that in odd dimensions  there exists a single deformation that describes a conformally massless spinor field. In even dimensions the minrep admits an infinite number of deformations that describe all the conformally massless higher spin fields. The minrep of the exceptional group $E_{6(-14)}$ with the maximal compact subgroup $SO(10)\times U(1)$ also admits infinitely many deformations \cite{GKN}.
 
 In the early days of supersymmetry, in search for an exceptional Lie superalgebra whose gauging would lead to a unified theory of all interactions including gravity,  the concept of generalized spacetimes coordinatized by Jordan algebras was introduced in \cite{Gunaydin:1975mp}. Jordan algebras $J_2^{\mathbb{A}}$ of $2\times 2$ Hermitian matrices over the four division algebras $\mathbb{A}=\mathbb{R},\mathbb{C},\mathbb{H},\mathbb{O}$ describe Minkowskian spacetimes in critical dimensions ($d=3,4,6,10$). Spacetimes defined by Jordan algebras  $J_3^{\mathbb{A}}$ of  $3\times 3$ Hermitian  matrices over $\mathbb{A}$  describe extensions of the Minkowskian spacetimes in critical dimensions by twistorial coordinates plus a single dilatonic coordinate\cite{Gunaydin:1975mp,Sierra:1986dx,Gunaydin:1992zh,Gunaydin:2005zz}. They are causal spacetimes whose conformal groups are $Sp(6,\mathbb{R}), SU(3,3), SO^*(12)$ and $E_{7(-25)}$.\footnote{ That the spacetimes defined by Euclidean Jordan algebras are causal was proven in \cite{Mack:2004pv}.}
The minimal unitary representations of the conformal groups of these spacetimes  were given in \cite{Gunaydin:2004md,Gunaydin:2005zz}.
These minreps are unitary lowest weight representations and their lowest energy irreps are singlets under the compact forms of their  Lorentz groups which are $SU(3)$, $SU(3)\times SU(3)$, $SU(6)$ and $E_6$.  However they turn out to be infinitely reducible when restricted to their Minkowskian conformal subgroups
$SO(d,2)\times SO(2,1)$ ($d=3,4,6,10$). In this paper we will study the decomposition of the minrep of the conformal group $E_{7(-25)}$ of the spacetime defined by the exceptional Jordan algebra $J_3^{\mathbb{O}}$  into representations of its subgroup $SO(10,2)\times SU(1,1)$. As a generalized conformal group   $E_{7(-25)}$ has a nonlinear action on the exceptional spacetime of dimension 27 defined by  $J_3^{\mathbb{O}}$. However the  minrep  of $E_{7(-25)}$ is realized over an Hilbert space of functions in 17 variables which correspond to 16 twistorial coordinates transforming in the spinor representation of $SO(10)$ plus the singlet dilatonic coordinate. We also give the decomposition of the minrep of $E_{7(-25)}$ with respect to its subgroup $SO^*(12)\times SU(2)$. The group $SO^*(12)$ is the conformal group of the spacetime coordinatized by the Jordan algebra $J_3^{\mathbb{H}}$ of $3\times 3$ Hermitian matrices over the division algebra $\mathbb{H}$ of quaternions.

The group $E_{7(-25)}$ is also the U-duality group of the $N=2$ Maxwell-Einstein supergravity defined by the exceptional Jordan algebra. We discuss the possible applications  of our results to the exceptional supergravity. 

The plan of the paper is as follows. In section 2 we review the extension of Minkowskian spacetimes in critical dimensions by twistorial and dilatonic coordinates. Sections 3 and 4 gives a review of the quasiconformal approach to minimal unitary representations  and its application to $E_{7(-25)}$. 
In section 5 we study the unitary representations of $SO(10,2)$ subgroup of $E_{7(-25)}$ constructed over a Fock space of oscillators transforming in a spinor representation of $SO(10)$. This is followed by the study of representations of a distinguished $SU(1,1)_K$ subgroup as symmetry of conformal quantum mechanics over the dilatonic coordinate whose coupling constant is given by the Casimir of $SO(10,2)$ subgroup. In section 7 we study the decomposition of Lie algebra of $E_{7(-25)}$ with respect to the Lie algebra of its maximal compact subgroup $E_6\times U(1)$. In section 8 we show that the minrep of $E_{7(-25)}$  decomposes into infinitely many representations of $SO(10,2)$ corresponding to massless conformal fields that transform as symmetric tensors in spinor indices subject to certain $SO(9,1)$ covariant constraints. Each irrep  of $SO(10,2)$ comes with infinite multiplicity that form an irrep of a distinguished $SU(1,1)_K$ subgroup. Conformal fields transforming as even symmetric tensors in spinor indices   describe bosons and odd symmetric tensors describe fermions. The noncompact generators of $E_{7(-25)}$ that transform as a spinor (16) of $SO(10)$ and its conjugate $\bar{16}$ act as  "bosonic supersymmetry" generators interpolating  between representations describing bosonic  and fermionic conformal fields. Hence the minimal unitary representation of $E_{7(-25)}$ provides an example of a bosonic symmetry whose spectrum exhibits spacetime "supersymmetry".  

In section 9 we show that the minrep of $E_{7(-25)}$ admits  deformations labelled by the Casimir operator  $SO(10,2)$ that is realized in terms of 12 bosonic oscillators. These deformations describe massive as well as massless conformal fields in ten dimensions and detailed analysis of these deformations will be given elsewhere. 

We also study the decomposition of the minrep of $E_{7(-25)}$ into representations of its subgroup $SO^*(12)\times SU(2)$ in section 10. $SO^*(12)$ is the conformal group of the spacetime coordinatized by the Jordan algebra $J_3^{\mathbb{H}}$
and its Lorentz group is $SU^*(6)$ with the maximal compact subgroup $USp(6)$.   
The minrep of $E_{7(-25)}$ decomposes into an infinite set of irreps of $SO^*(12)$. Each irrep of $SO^*(12)$ comes  with finite multiplicity and transform  in a representation of $SU(2)$.

 The group $E_{7(-25)}$ is also the $4d$ U-duality  group of the exceptional Maxwell-Einstein supergravity defined by the exceptional Jordan algebra\cite{Gunaydin:1983rk,Gunaydin:1983bi}. 
 In section 11 we show that the composite scenario proposed for the exceptional supergravity\cite{Gunaydin:1983rk,Gunaydin:1984be}  can, in principle, accommodate infinitely many chiral families of leptons and quarks transforming in the spinor 16 representation of the grand unified group $SO(10)$ with ever increasing masses. Again in section 11 we discuss possible applications of our results to the proposal that  $E_{7(-25)}$ must act as spectrum generating symmetry group of the $5d$ exceptional supergravity. 
 
In Appendix A we review the minimal unitary representation  of $E_{8(-24)}$ given in \cite{Gunaydin:2004md,Gunaydin:2005zz}. In Appendix B we give the construction of the irreps of $SO(10)$ subgroup of $SO(10,2)$ in terms of bosonic oscillators in a $U(5)$ covariant basis. In Appendix C we study the Lie algebra of the  compact $E_6$ subgroup of $E_{7(-25)}$.   Appendix D  discusses the truncation of the results obtained for $E_{7(-25)}$ to subgroups $SO^*(12), SU(3)\times SU(3)$ and $Sp(6,\mathbb{R})$ and their relation to conformal symmetry in lower critical spacetime dimensions $d=6,4$ and $d=3$, respectively.

\section{ Extensions of Minkowskian spacetimes in critical dimensions by twistorial coordinates
\label{sec:SpaceTime-Ext}}

In  generalized spacetimes defined by Jordan algebras $J$ one identifies the automorphism group 
of $J$ with the rotation group, the reduced structure group of $J$ with the Lorentz group and the linear fractional group of $J$ with the conformal group \cite{Gunaydin:1975mp,Gunaydin:1979df,Gunaydin:1992zh}. For Euclidean ( formally real) Jordan algebras  the conformal groups of $J$ admit positive energy unitary representations \cite{Gunaydin:1981yq,Gunaydin:1999jb} and the corresponding spacetimes have been shown to be causal\cite{Mack:2004pv}. Degree two Euclidean  Jordan  algebras describe ordinary Minkowskian space-times and their norms are given by quadratic forms $Q$ of Minkowskian signature 
\bea
Q(x) = \eta^{\mu\nu} x_\mu x_\nu \quad, ,\quad \mu ,\nu, ..=0,1,...,d-1 
\eea
where $ \eta^{\mu\nu}= \text{Diag}  ( 1,-1,-1,..,-1) $ and we shall label them as $\Gamma(Q)$. 

There exist an infinite family of reducible Jordan algebras of degree three which are a direct sum of simple rank two Jordan algebras $\Gamma(Q)$ and a one dimensional Jordan algebra $\mathbb{R}$ 
\bea
J_3(Q) = \mathbb{R} \oplus \Gamma(Q)
\eea
which we shall refer to as generic Jordan family of degree three. They describe extensions of Minkowskian spacetimes by a dilatonic coordinate which we label as $\rho$. The cubic norm has the form
\begin{equation}
    \mathcal{V}\left(\rho,x_\mu \right) = \sqrt{2} \rho \, x_\mu x_\nu
    \eta^{\mu\nu}
\end{equation}
The rotation  group of this spacetime is the automorphism group $SO(d-1)$ and the Lorentz group of
 is the reduced structure group 
\begin{equation}
  \mathrm{SO}(d-1,1) \times \mathrm{SO}(1,1)
\end{equation}
Under $SO(1,1)$ action the coordinates transform as 
\begin{equation}
   SO(1,1): \begin{aligned} \rho &\Rightarrow e^{2\lambda} \rho \\
  x_{\mu} & \Rightarrow e^{-\lambda} x_{\mu} \end{aligned}
\end{equation}
The simple Euclidean Jordan algebras of degree two and dimensions $3,4,6 $ and $10$ are isomophic to Jordan algebras defined by $2\times 2$ Hermitian matrices over the four division algebras $\mathbb{R}, \mathbb{C} ,\mathbb{H} ,\mathbb{O}$. They correspond to the critical Minkowskian space-times for the existence of supersymmetric Yang-Mills theories. 

In addition to the  infinite family of reducible Euclidean Jordan algebras   there exist four simple Jordan algebras of degree three defined by $3\times 3$ Hermitian  matrices over the four division algebras which are denoted as $J_3^{\mathbb{A}}$ for $\mathbb{A}= \mathbb{R}, \mathbb{C} ,\mathbb{H} ,\mathbb{O}$.
They describe extensions of Minkowskian space-times in critical dimensions $d=3,4,6,10$ with twistorial coordinates $\xi^\alpha$ transforming in a spinor representation of the Lorentz group and an extra dilatonic coordinate $\rho$. Using the conventions of \cite{Sierra:1986dx} we have
\begin{equation}
\begin{split}
   J_3^\mathbb{R}  &\Longleftrightarrow \left(\rho, x_\mu, \xi^{\alpha} \right) \quad \mu =0,1,2 \quad \alpha =1,2 \\
   J_3^\mathbb{C}  &\Longleftrightarrow \left(\rho, x_\mu, \xi^{\alpha} \right) \quad \mu =0,1,2,3 \quad \alpha =1,2,3,4 \\
   J_3^\mathbb{H}  &\Longleftrightarrow \left(\rho, x_\mu, \xi^{\alpha} \right) \quad \mu =0,\ldots,5 \quad \alpha =1,\ldots,8 \\
   J_3^\mathbb{O}  &\Longleftrightarrow \left(\rho, x_\mu, \xi^{\alpha} \right) \quad \mu =0,\ldots,9 \quad \alpha =1,\ldots,16
\end{split}
\end{equation}
The Minkowskian coordinates $x_\mu$ are represented by the Hermitian matrices $J_2^{\mathbb{A}}$ and twistorial coordinates $\xi^\alpha$ by $2 \times 1$   matrix $\xi$
over $\mathbb{A}=\mathbb{R}, \mathbb{C},\mathbb{H},\mathbb{O}$. Hence the elements of $J_3^{\mathbb{A}}$ decompose as 
\[
J_3^{\mathbb{A}} = 
\left(\begin{array}{c|c} J_2^{\mathbb{A}}  & \xi   \\ \hline   \bar{\xi}  & \rho  \end{array}\right) \]
The cubic norm $\mathcal{V}(J)$ of an element $J$ of $J_3^{\mathbb{A}}$ is given by \cite{Sierra:1986dx}
\begin{equation}
  \mathcal{V}\left(\rho, x_\mu, \xi^{\alpha}\right) = \sqrt{2} \rho x_\mu x_\nu \eta^{\mu\nu} + x^\mu \Bar{\xi} \gamma_\mu \xi
\end{equation}
where $\gamma_\mu$ are the sigma matrices in the corresponding critical dimension. 
The Lorentz groups $Lor(J)$ of the space-times over $J_3^\mathbb{A}$ are given by the invariance groups of their cubic norm forms which are
\begin{equation}
\begin{split}
 Lor(J_3^{\mathbb{R}} ) & :   \mathrm{SL}\left(3, \mathbb{R}\right) \cr
 Lor( J_3^{\mathbb{C}} ) &: \mathrm{SL}\left(3, \mathbb{C}\right) \cr
 Lor(  J_3^{\mathbb{H}} ) & :  \mathrm{SU}^\ast\left(6\right) \cr
 Lor(  J_3^{\mathbb{O}}) & : \mathrm{E}_{6(-26)}
   \end{split} 
\end{equation}
The Lorentz groups of the spacetimes defined by their reducible subalgebras are
\begin{equation}
\begin{split}
    Lor( \mathbb{R} \oplus J_2^\mathbb{R} ) &: \mathrm{SO}(1,1) \times \mathrm{SO}\left(2,1\right) \subset \mathrm{SL}\left(3, \mathbb{R}\right) \cr
   Lor(\mathbb{R} \oplus J_2^\mathbb{C} ) &: \mathrm{SO}(1,1) \times \mathrm{SO}\left(3,1\right) \subset \mathrm{SL}\left(3, \mathbb{C}\right) \cr
   Lor(\mathbb{R} \oplus J_2^\mathbb{H} ) &: \mathrm{SO}(1,1) \times \mathrm{SO}\left(5,1\right) \subset \mathrm{SU}^\ast\left(6\right) \cr
   Lor(\mathbb{R} \oplus J_2^\mathbb{O} ) &: \mathrm{SO}(1,1) \times \mathrm{SO}\left(9,1\right) \subset \mathrm{E}_{6(-26)}
\end{split}
\end{equation}
In \cite{Sierra:1986dx} it was shown that the adjoint identity satisfied by the elements of the Jordan algebras $J_3^\mathbb{A}$ imply the Fierz identities for the existence of super Yang-Mills theories in the critical Minkowskian spacetime dimensions $d=3,4,6,10$.

The conformal groups  $Conf(J)$ of these spacetimes are 
\begin{equation}
\begin{split}
Conf(J_3^{\mathbb{R}} ) & :   \mathrm{Sp}\left(6, \mathbb{R}\right) \cr
Conf( J_3^{\mathbb{C}} ) &: \mathrm{SU}\left(3,3 \right) \cr
Conf(  J_3^{\mathbb{H}} ) & :  \mathrm{SO}^\ast\left(12\right) \cr
Conf(  J_3^{\mathbb{O}}) & : \mathrm{E}_{7(-25)}
   \end{split} 
\end{equation}

\begin{equation}
\begin{split}
   Conf( \mathbb{R} \oplus J_2^\mathbb{R} ) &: \mathrm{SO}(2,1) \times \mathrm{SO}\left(3,2\right) \subset \mathrm{Sp}\left(6, \mathbb{R}\right) \cr
  Conf(\mathbb{R} \oplus J_2^\mathbb{C} ) &: \mathrm{SO}(2,1) \times \mathrm{SO}\left(4,2\right) \subset \mathrm{SU}\left(3,3 \right) \cr
  Conf(\mathbb{R} \oplus J_2^\mathbb{H} ) &: \mathrm{SO}(2,1) \times \mathrm{SO}\left(6,2\right) \subset \mathrm{SO}^\ast\left(12\right) \cr
  Conf(\mathbb{R} \oplus J_2^\mathbb{O} ) &: \mathrm{SO}(2,1) \times \mathrm{SO}\left(10,2\right) \subset \mathrm{E}_{7(-25)}
\end{split}
\end{equation}

Conformal groups $Conf(J)$ of a Jordan algebra leaves invariant  light-like separations  with respect to their norm. Hence for  degree three Jordan algebras $J$ the group $Conf(J)$ is referred to as the invariance group of cubic light-cone defined over  $J$. It was shown in \cite{Gunaydin:2000xr} that the nonlinear action of $Conf(J)$ on $J$ admits an extension to the nonlinear action of a larger  group, called the quasiconformal group,  $QConf(J)$ on a space coordinatized by the Freudenthal triple system $\mathcal{F}(J)$  defined over $J$ extended by a singlet cocycle coordinate $x$. This quasiconformal group action leaves invariant light-like separations with respect to a quartic norm and hence it is referred to as the invariance group of a
quartic light-cone. The quasiconformal groups associated with Euclidean Jordan algebras of degree three are:
\begin{equation}
\begin{split}
QConf(J_3^{\mathbb{R}} ) & :   \mathrm{F}_{4(4)} \cr
QConf( J_3^{\mathbb{C}} ) &: \mathrm{E}_{6(2)} \cr
QConf(  J_3^{\mathbb{H}} ) & :  \mathrm{E}_{7(-5)} \cr
QConf(  J_3^{\mathbb{O}}) & : \mathrm{E}_{8(-24)} \cr
QConf( \mathbb{R} + \Gamma(Q)) &: \mathrm{SO}(d,4) 
   \end{split} 
\end{equation}

Quasiconformal realizations extend to all simple Lie groups of dimension greater than 3. Most importantly the quantization of this geometric realization of non-compact groups leads directly to their minimal unitary representations \cite{Gunaydin:2001bt,Gunaydin:2004md,Gunaydin:2005zz,Gunaydin:2006vz}. In the next section we summarize this construction briefly.

\section{Quasiconformal Group Approach to Minimal Unitary Representations of Non-compact Groups}

Motivated by the work of physicists on spectrum generating algebras Joseph introduced the concept of a minimal unitary representation of a noncompact group 
 $G$ and  determined the minimal dimensions of noncompact groups \cite{Joseph:1974}.  These dimensions turn out to be given simply  by considering a particular 5-graded decomposition of   non-compact Lie algebras $\mathfrak{g}$ of the form\cite{Gunaydin:2001bt,Gunaydin:2006vz}
\begin{equation}
 \mathfrak{g}= \mathfrak{g}^{-2}  \oplus  \mathfrak{g}^{-1}  \oplus
  \left(  \mathfrak{g}^{0}  \oplus \Delta \right)
  \oplus \mathfrak{g}^{+1}  \oplus  \mathfrak{g}^{+2}
\end{equation}
where  grade $\pm 2$ dimensional subspaces $\mathfrak{g}^{\pm 2}$ are one dimensional. The generator  $\Delta$ that determines the 5-grading forms  a distinguished  $\mathfrak{sl}(2,\mathbb{R})$  subalgebra together with the generators  $\mathfrak{g}^{\pm 2}$  
 \[\mathfrak{sl}(2,\mathbb{R})=\mathfrak{g}^{-2} \oplus \Delta \oplus \mathfrak{g}^{+2}\]

The  dimension $\ell$ of the minimal unitary representation of $G$  is related to the dimension of grade $+1$ subspace of $ \mathfrak{g}$: 
\begin{equation}
   \ell=\frac{1}{2} \dim \left(\mathfrak{g}^{+1}\right) +1
\end{equation}

Denoting  the subgroup generated by the  subalgebra
$\mathfrak{g}^0$ as $H$, then the non-compact coset space
\begin{equation}
 \frac{G}{H\times SL(2,\mathbb{R})}
\end{equation}
 is a para-quaternionic symmetric space. The
 minimal unitary representations of all such
non-compact groups were constructed in a unified manner in \cite{Gunaydin:2006vz} by quantizing  their quasiconformal
realizations  \cite{Gunaydin:2000xr} which generalized the earlier constructions of the minimal unitary representations of 
exceptional groups ($F_4$, $E_6$, $E_7$, $E_8$) and of $SO(n,4)$ given in
\cite{Gunaydin:2001bt,Gunaydin:2004md,Gunaydin:2005zz}.
Simple
noncompact groups $G$ of this type and their subgroups $H$ are given in Table \ref{listG}. 
\begin{table} 
\begin{center}
\begin{tabular}{|c|c|}
\hline $G$ & $H$ \\ \hline
$SU(m,n)$ & $U(m-1,n-1)$ \\ \hline
$SL(n,\mathbb{R})$ & $GL(n-2,\mathbb{R})$ \\ \hline
$SO(n,m)$ & $SO(n-2,m-2)\times SU(1,1)$ \\ \hline
$SO^*(2n)$ & $SO^*(2 n-4)\times SU(2)$ \\ \hline
$Sp(2n,\mathbb{R})$ & $Sp(2n-2,\mathbb{R})$ \\ \hline
$E_{6(6)}$ & $SL(6,\mathbb{R})$ \\ \hline
$E_{6(2)}$ & $SU(3,3)$ \\ \hline
$E_{6(-14)}$ & $SU(5,1)$ \\ \hline
$E_{7(7)}$ & $SO(6,6)$ \\ \hline
$E_{7(-5)} $ & $SO^*(12)$ \\ \hline
$E_{7(-25)}$ & $SO(10,2) $ \\ \hline
$E_{8(8)}$ & $E_{7(7)}$ \\ \hline
$E_{8(-24)} $ & $E_{7(-25)} $ \\ \hline
$F_{4(4)}$ & $Sp(6,\mathbb{R})$ \\ \hline
$G_{2(2)}$ & $SU(1,1)$ \\ \hline
\end{tabular}
\end{center}
\caption{Simple noncompact groups $G$ and their subgroups $H$ such that $\frac{G}{H\times SL(2,\mathbb{R})} $ is a para-quaternionic symmetric space. \label{listG} }
\end{table}

Let $J^a$ be the  generators of the subgroup 
$H$ of dimension $D$ with the commutation relations
\begin{subequations}\label{eq:alg}
\begin{equation}
   \left[ J^a \,, J^b \right] = {f^{ab}}_c J^c
\end{equation}
where $a,b,...=1,...D$. The    subspaces $\mathfrak{g}^{- 1}$ and $\mathfrak{g}^{+ 1}$ with generators $E^\alpha$ and $F^\alpha$ , respectively,  form a symplectic representation   $\rho$ under the action of $H$
\begin{equation}
   \left[ J^a \,, E^{\alpha} \right] = {\left(\lambda^{a}\right)^\alpha}_\beta E^\beta
\qquad
   \left[ J^a \,, F^{\alpha} \right] =  {\left(\lambda^{a}\right)^\alpha}_\beta F^\beta
\end{equation}
where  $ \alpha, \beta, ..= 1,..,N= \dim (\rho)$. They close into the generators $E$ and $F$  in grade $-2$ and $+2$ subspaces  and form Heisenberg algebras of dimension $N+1$, respectively: 
\begin{equation}
   \left[ E^\alpha \,, E^\beta  \right] = 2 \Omega^{\alpha\beta} E
\end{equation}

\begin{equation}
   \left[ F^\alpha \,, F^\beta  \right] = 2 \Omega^{\alpha\beta} F
\end{equation}
where  $\Omega^{\alpha\beta}$ is the symplectic invariant ``metric'' of
the representation $\rho$.  The
remaining nonvanishing commuttators of the generators  of $g$ are simply of the form
\begin{equation}
\begin{aligned}
  F^\alpha &= \left[ E^\alpha \,, F \right] \cr
  E^\alpha &= \left[ E \,, F^\alpha \right] \cr
  \left[E^{\alpha} , F^{\beta}\right] &= - \Omega^{\alpha\beta} \Delta + \epsilon \lambda_a^{\alpha\beta} J^a
\end{aligned}
\qquad
\quad
\begin{aligned}
\left[\Delta, E^{\alpha} \right] &= - E^{\alpha} \cr
\left[\Delta, F^{\alpha} \right] &= F^{\alpha} \cr
\left[\Delta, E\right] &= -2 E \cr
\left[\Delta, F \right] &= 2 F
\end{aligned}
\end{equation}
\end{subequations}
where 
$\epsilon$ is a  parameter that depends on $g$.

 \section{The minimal unitary realization of  the exceptional Lie group $E_{7(-25)}$ }

\label{minrep-e7-25}

As a  conformal group of the exceptional Jordan algebra $J_3^{\mathbb{O}}$ the Lie algebra of  $\mathrm{E}_{7(-25)}$ admits a  three-graded decomposition with respect to the  generator of dilatation group $SO(1,1)$:
\begin{equation}
\mathfrak{e}_{7(-25)} = \overline{27} \oplus (\mathfrak{e}_{6(-26)}
\oplus \mathfrak{so}(1,1) ) \oplus 27
\end{equation}
with ($\bar{27})$ denotes the grade  $-1$  generators and $27$ denotes the grade $+1$  generators corresponding to translations and special conformal transformations, respectively. The grade zero generators that commute with $SO(1,1)$ form the Lie algebra $ \mathfrak{e}_{6(-26)}$ of the Lorentz group $E_{6(-26)}$  of the exceptional space-time defined by $J_3^{\mathbb{O}}$.

On the other hand as a quasiconformal Lie algebra $\mathfrak{e}_{7(-25)}$ admits a natural 5-graded decomposition with respect to its subalgebra $\mathfrak{so}(10,2)\oplus \mathfrak{so}(1,1)$
\begin{equation}
 \mathfrak{e}_{7(-25)} \mspace{5mu} = \mspace{5mu}
\begin{array}{ccccccccc}
   \mathfrak{g}^{-2} & \oplus & \mathfrak{g}^{-1} & \oplus & \mathfrak{g}^{0} & \oplus &
   \mathfrak{g}^{+1} & \oplus & \mathfrak{g}^{+2} \\
   \mathbf{1} & \oplus & \mathbf{32} & \oplus & \left( \mathfrak{so}(10,2)\oplus \Delta \right) & \oplus &
   \mathbf{32} & \oplus & \mathbf{1}
\end{array}
\end{equation}
and its minimal unitary representation can be obtained by quantization of its geometric realization as a quasiconformal group. It was first obtained by truncation of the quasiconformal Lie algebra of $E_{8(-24)}$ \cite{Gunaydin:2004md,Gunaydin:2005zz}. Minimal unitary representation of $E_{8(-24)}$ ,which we review in  Appendix A, is realized over the tensor product of the Fock space of 28 bosonic oscillators $\Tilde{Z}_{ab}=- \Tilde{Z}_{ba}\, \, (Z^{ab}=-Z^{ba})$  ($a,b,..=1,...,8$) with the state space of a conformal quantum mechanics spanned by wave functions that depend  on the singlet coordinate $x$. 
Truncation of the minimal unitary realization of  $\mathfrak{e}_{7(-25)} $ from that of $\mathfrak{e}_{8(-24)} $ is achieved  by  setting $Z^{a8} = Z^{8a} = 0$ where $a \not= 7$, as well as $Z^{a7} = Z^{7a} = 0$ for $a \not= 8$: 
Hence the corresponding Hilbert space of the minrep of $E_{7(-25)}$ is spanned by  the tensor product of the Fock space of 16 bosonic oscillators $Z^{ab}=-Z^{ba} (a,b=1,\cdots, 6 ), Z^{78}$  and the Hilbert space of conformal quantum mechanics.

The Lie subalgebra $\mathfrak{so}(10,2)$ has a three-grading with respect to its $\mathfrak{u}(5,1)$ subalgebra
\eq
 \mathfrak{so}(10,2) =  \mathfrak{f}^{-1}  \oplus  \mathfrak{f}^{0} \oplus 
   \mathfrak{f}^{+1}
 \en
where $\mathfrak{f}^{0}=\mathfrak{su}(5,1)+\mathfrak{u}(1) $. The  generators of $\mathfrak{f}^{0}=\mathfrak{su}(5,1)+\mathfrak{u}(1) $ are given by the bilinears

\begin{equation}
    {J^{a}}_{b} = 2 Z^{ac} \Tilde{Z}_{bc} - \frac{1}{3} {\delta^{a}}_{b}
    Z^{dc} \Tilde{Z}_{dc}
\end{equation}
and 
\eq
 U = \frac{3}{2} \left( Z^{78} \Tilde{Z}_{78} + \Tilde{Z}_{78} Z^{78} \right) - \frac{1}{4} \left( Z^{ab}
   \Tilde{Z}_{ab} + \Tilde{Z}_{ab} Z^{ab} \right) 
   \en
where the indices $a,b,..$ now run from 1 to 6. The other generators of $\mathfrak{so}(10,2)$ belonging to grade $\pm 1$ subspaces $\mathfrak{f}^{\pm 1}$  are as follows:
\begin{equation}
\begin{aligned} 
  J^+_{ab} &= - \frac{1}{6} \Tilde{Z}_{ab} \Tilde{Z}_{78} + \frac{1}{48}
   \epsilon_{abcdef} Z^{cd} Z^{ef} \cr
  J_-^{ab} &= \frac{1}{6} Z^{ab} Z^{78} - \frac{1}{48}
  \epsilon^{abcdef} \Tilde{Z}_{cd} \Tilde{Z}_{ef}
\end{aligned}
\end{equation}
where $a,b,..=1,...,6$. The oscillators $Z^{ab}$ ($\tilde{Z}_{ab} $) satisfy the canonical commutation relations 
\bea
[\tilde{Z}_{ab}, Z^{cd} ] = \frac{1}{2} (\delta_a^c \delta_b^d -\delta_b^c  \delta_a^d) :=\delta^{cd}_{ab} 
\eea
\bea
[\tilde{Z}_{78}, Z^{78}] =1/2 
\eea
and  the Hermiticity conditions
\begin{equation}
    \left(Z^{ab}\right)^\dagger  =\tilde{\eta}^{ac} \tilde{\eta}^{bd} \Tilde{Z}_{cd} \quad , \, \tilde{Z}_{78}^\dagger = - Z^{78} 
\end{equation}
where $\tilde{\eta}_{ab} = \mathop\mathrm{diag}\left(+1,+1,+1,+1,+1,-1\right)$ .
They imply 
\begin{equation}
   \left( {J^{a}}_{b} \right)^\dagger = \tilde{\eta}^{ac} \tilde{\eta}_{bd} {J^{d}}_{c}
  \qquad U^\dagger = U \qquad \left( J^+_{ab} \right)^\dagger = J_-^{cd} \tilde{\eta}_{ac}
   \tilde{\eta}_{bd}
\end{equation}

Generators of $SO(10,2)$ satisfy  the commutation relations
\begin{equation}
\begin{aligned}
   \left[ {J^{a}}_{b} ,\, {J^{c}}_{d}  \right] &= {\delta^{c}}_{b} {J^{a}}_{d}
   -  {\delta^{a}}_{d} {J^{c}}_{b}\cr
   \left[ {J^{a}}_{b} ,\, J_-^{cd} \right] &= {\delta^{c}}_{b} J_-^{ad}
  + {\delta^{d}}_{b} J_-^{ca} - \frac{1}{3} {\delta^{a}}_{b} J_-^{cd} \cr
   \left[ {J^{a}}_{b} ,\, J^+_{cd} \right] &= -{\delta^{a}}_{c} J^+_{bd}
  - {\delta^{a}}_{d} J^+_{cb} + \frac{1}{3} {\delta^{a}}_{b} J^+_{cd}\cr
   \left[ U ,\, J^+_{cd} \right] &= - J^+_{cd} \qquad ,
   \left[ U ,\, J_-^{cd} \right] = + J_-^{cd}  \qquad ,
   \left[ U ,\, {J^{c}}_{d} \right] = 0 \qquad \cr
   \left[ J^+_{ab} , J_-^{cd} \right] &= \frac{1}{144} ( \delta^c_b J^d_a - \delta^c_a J^d_b -\delta^d_b J^c_a + \delta^d_a J^c_b )- \frac{1}{108} \delta^{cd}_{ab} \,  U
\end{aligned}
\end{equation}
The creation operators $\Tilde{Z}_{78}, \, Z^{mn} $ and $\Tilde{Z}_{m6} \, (m,n=1,\cdots , 5)$ transform in the spinor representation $16$ of $SO(10)$ which decomposes as $ 1 + 10 + \bar{5}$ under the $SU(5)$ subgroup. The corresponding annihilation operators transform in the conjugate spinor $\bar{16}$ of $SO(10)$ and together they form the $32$ dimensional Majorana-Weyl spinor of $SO(10,2)$. 

The quadratic Casimir of $SO(10,2)$  is given by
\begin{equation}
   \mathcal{C}_2\left(\mathfrak{so}\left(10,2\right)\right) =
    \frac{1}{6} {J^{a}}_{b} {J^{b}}_{a} + \frac{1}{9} U^2 + 12 \left( J_-^{ab}
   J^+_{ab} +    J^+_{ab} J_-^{ab}  \right) = I_4 - \frac{99}{16}
\end{equation}
where  $I_4$ is the quartic invariant of $SO(10,2)$ that descends from the quartic invariant of $E_{7(-25)}$ given in equation \ref{quartic-e7} under truncation.

To define the grade $\pm 2$ and $\pm 1$ generators we introduce one dimensional quantum mechanical coordinate $x$ and momentum operator $p$ that satisfy the commutation relation
\bea
[x,p]=i
\eea 
Grade -2 generator is simply
\bea 
M= \frac{1}{2} x^2 
\eea
and the grade $-1$ generators are given by
\bea
M^{ab}= x Z^{ab}  \quad,  \quad  M^{78} = x Z^{78} \\ \nn
\Tilde{M}_{ab} = x \Tilde{Z}_{ab} \quad , \quad  \Tilde{M}_{78} = x \Tilde{Z}_{78} 
\eea
where $a,b,..=1,...,6$. The grade $-1$ generators $(M^{ab} , M^{78} )$ and $(\Tilde{M}_{ab}, \Tilde{Z}_{78} )$ transform in  spinor representation $16$ and $\bar{16}$ of $SO(10)$ and together they form a Majorana-Weyl spinor of  $SO(10,2)$. 

Grade $-1$ generators close into the grade $-2$ generator and form an Heisenberg algebra
\bea
[\Tilde{M}_{ab} , M^{cd} ] =(\delta_a^c \delta_b^d -\delta_b^c 
\delta_a^d ) M
\eea

The grade $+2$ generator is given by
\bea
K=\frac{p^2}{2}  +\frac{2}{x^2} \left( C_2 + \frac{99}{16} \right)
\eea
where $ C_2$ is the Casimir of grade zero subalgebra $\mathfrak{so}(10,2) $. The 
 grade $+1$ generators are given by
\bea
K^{ab} = i [ M^{ab} , K]  \quad , \quad \Tilde{K}_{ab} = i [ \Tilde{M}_{ab} , K ] 
\eea
and 
\bea
K^{78} = i [M^{78} , K] \quad , \quad \Tilde{K}_{78} = i [ \Tilde{M}_{78}, K ] 
\eea
They close into  the grade $+2$ generator $K$  and together they form an Heisenberg algebra as well.

Commutation relations of  generators in $\mathfrak{g}^0$
with those of $\mathfrak{g}^{-1}$ read as
\begin{equation*}
\begin{split}
   \left[ {J^a}_b \,, M^{cd} \right] &= \delta^{c}_b M^{ad} + \delta^d_b M^{ca} - \frac{1}{3} \delta^a_b M^{cd} \qquad
   \left[ {J^a}_b \,, \Tilde{M}_{cd} \right] =  -{\delta^{a}_c} \Tilde{M}_{bd} -{\delta^{a}_d} \Tilde{M}_{cb} + \frac{1}{3} \delta^a_b \Tilde{M}_{cd} \\
   \left[ J^{ab}_- \,, M^{cd} \right] &= - \frac{1}{24} \epsilon^{abcdef} \Tilde{M}_{ef} \qquad
     \left[ J^+_{ab} \,, M^{cd} \right] =  - \frac{1}{6} \delta^{cd}_{ab} \Tilde{M}_{78} \\
    \left[ J^+_{ab} \,, \Tilde{M}_{cd} \right] &= - \frac{1}{24} \epsilon_{abcdef} M^{ef} \qquad
    \left[ J^{ab}_- \,, \Tilde{M}_{cd} \right] = - \frac{1}{6} \delta^{ab}_{cd} M^{78}
\end{split}
\end{equation*}
\begin{equation*}
\begin{split}
    \left[ U \,, M^{ab} \right] &= -\frac{1}{2} M^{ab} \qquad
       \left[ U \,, \Tilde{M}_{ab} \right] = + \frac{1}{2} \Tilde{M}_{ab} \\
    \left[ J^{ab}_- \,, M^{78} \right] &= 0 \qquad
     \left[ J^+_{ab} \,, M^{78} \right] = - \frac{1}{12} \Tilde{M}_{ab} \qquad
     \left[ U \,, M^{78} \right] =  \frac{3}{2} M^{78}  \\
     \left[J^+_{ab} \,, \Tilde{M}_{78} \right] & = 0 \qquad
     \left[ J_-^{ab} \,, \Tilde{M}_{78} \right] = - \frac{1}{12} M^{ab} \qquad
     \left[ U \,, \Tilde{M}_{78} \right] = - \frac{3}{2} \Tilde{M}_{78}
\end{split}
\end{equation*}
Five grading of $\mathfrak{e}_{7(-25)}$ is determined by the $SO(1,1)$ generator $\Delta=-\frac{i}{2} ( x p + p x) $ where $p$ is the momentum conjugate to $x$. 
The grade $\pm 2$ generators close into $\Delta$ and form an $SU(1,1)_K $ subalgebra
 \begin{equation}
   \left[ M, K \right] = - \Delta \qquad
   \left[ \Delta \,, M \right] = - 2 M \qquad
   \left[ \Delta \,, K \right] = + 2 K \,.
\end{equation}

Generators in $\mathfrak{g}^{-1}$ and $\mathfrak{g}^{+1}$ close into generators in $\mathfrak{g}^{0}$ 
\begin{equation}
\begin{split}
 \left[ M^{ab} \,, K^{cd} \right] &= -6 i \epsilon^{abcdef} J^+_{ef} \\
 \left[ M^{ab} \,, \Tilde{K}_{cd} \right] &=  i \delta^{ab}_{cd} \left(2/3 \, U  +\Delta \right) - 4 i {\delta^{[a}}_{[c} {J^{b]}}_{d]} \\
 \left[ M^{ab} \,, K^{78} \right] &= -12 i J^{ab}_- \qquad
 \left[ M^{ab} \,, \Tilde{K}_{78} \right] = 0 \\
\end{split}
\end{equation}
\begin{equation}
\begin{split}
  \left[ \Tilde{M}_{ab} \,, K^{cd} \right] &= - i \delta^{cd}_{ab} \left( -2/3 \, U + \Delta \right) + 4 i {\delta^{[c}}_{[a} {J^{d]}}_{b]} \\
  \left[ \Tilde{M}_{ab} \,, \Tilde{K}_{cd} \right] &= + 6 i \epsilon_{abcdef} J^{ef}_- \\
  \left[ \Tilde{M}_{ab} \,, \Tilde{K}_{78} \right] &= +12 i J_{ab}^+\qquad
    \left[ \Tilde{M}_{ab} \,, K^{78} \right] = 0
\end{split}
\end{equation}
\begin{equation}
\begin{split}
  \left[ M^{78} \,, K^{ab} \right] &= - 12 i J^{ab}_- \quad
  \left[ M^{78} \,, \Tilde{K}_{78} \right] = i \left( \frac{1}{2} \Delta -U \right)  \\
    \left[ M^{78} \,, \Tilde{K}_{ab} \right] &= 0 \qquad
  \left[ M^{78} \,, K^{78} \right] = 0 \\
  \left[ \Tilde{M}_{78}\,, K^{ab} \right] &= 0 \qquad
    \left[ \Tilde{M}_{78}\,, \Tilde{K}_{78} \right] = 0 \\
  \left[ \Tilde{M}_{78}\,, K^{78} \right] &= i \left( - \frac{1}{2} \Delta -  U \right) \quad
  \left[ \Tilde{M}_{78}\,, \Tilde{K}_{ab} \right] = + 12 i J_{ab}^+
\end{split}
\end{equation}
The  quadratic Casimir of the above minimal unitary realization of
$\mathfrak{e}_{7(-25)}$  reduces to a c-number 
\begin{equation}
\begin{split}
  \mathcal{C}_2 \left( \mathfrak{e}_{7(-25)} \right) &= \mathcal{C}_2 \left( \mathfrak{so}(10,2) \right) +
               \frac{1}{12} \Delta^2 + \frac{1}{6} \left( M K + KM \right) \\
                & - \frac{i}{12} \left( \Tilde{M}_{ab} K^{ab} +  K^{ab} \Tilde{M}_{ab}
                                          - \Tilde{K}_{ab}  M^{ab} - M^{ab} \Tilde{K}_{ab} \right)   \\
               & - \frac{i}{6} \left( \Tilde{M}_{78} K^{78} +  K^{78} \Tilde{M}_{78}
                                          - \Tilde{K}_{78} M^{78} - M^{78} \Tilde{K}_{78} \right) \\
               & = \left( I_4 -\frac{99}{16} \right) + \left( \frac{1}{3} I_4 - \frac{1}{16} \right) +
                   \left( - \frac{4}{3} I_4 - \frac{31}{4} \right) = - 14
\end{split}
\end{equation}

\section{ Unitary representations  of $SO(10,2)$ subgroup of $E_{7(-25)}$ and massless  conformal fields in ten dimensional Minkowskian spacetime} \label{covariant}  

In the previous section we gave a minimal unitary realization of $E_{7(-25)}$ using oscillators transforming covariantly with respect to $SU(5,1)$ subgroup. 
In this section, for  the purpose of obtaining the K-type decomposition of the irreps of $SO(10,2) $  that occur inside the minrep of $E_{7(-25)}$ we shall reformulate these results in an $SO(10)\times U(1)$  covariant basis. The construction of the relevant $SO(10)$  irreps in an $U(5)$ covariant basis is given in Appendix B. 



Let us denote the generators of $SO(10,2)$ as $S_{A,B}=-S_{B,A}$ where $A,B,...=1,...,12$. In terms of $U(5)$ covariant generators they are given by:
\bea
S_{m,n} & := &6( J^+_{mn} +J_-^{mn}) + i/2  (J^m_n-J^n_m)  \qquad m,n=1,..,5 \\
S_{m,n+5}=-S_{n+5,m} & := &6 i ( J^+_{mn} - J_-^{mn}) - 1/2  (J^m_n+J^n_m)+ 2/3 \delta_{m,n} U    \\
S_{m,11}=-S_{11,m} & :=& 6 i ( J^+_{m6} + J_-^{m6}) -1/2  (J^m_6-J^6_m)   \\
S_{m+5,11}=-S_{11,m+5} & :=& 6  ( J^+_{6m} - J_-^{6m}) + i/2  (J^m_6+J^6_m )   \\
S_{12,m}=-S_{m,12} & :=& -6 J^+_{m6}+6 J_-^{m6} -i/2 J^m_6 -i/2 J^6_m \\
S_{12,m+5}=-S_{m+5,12} & :=& -6 i  J^+_{m6}- 6i  J_-^{m6} -1/2 J^m_6 +1/2 J^6_m \qquad m=1,\cdots 5  \\
S_{11,12}&=&-S_{12,11} := - (J^6_6 +1/3 \, U) 
\eea
They satisfy the commutation relations 
\bea
[ S_{A,B} , S_{C,D} ] = - i \left( \theta_{AC} S_{B,D} + \theta_{BD} S_{A,C} - \theta_{BC} S_{A,D} - \theta_{AD} S_{B,C} \right)
\eea
where $\theta_{AB} = \mathrm{Diag} (1,1,1,1,1,1,1,1,1,1,-1,-1) $ ( $A,B,\cdots= 1,2, \cdots ,12$).
Lie algebra of $SO(10,2)$ admits a compact three grading given by  the generator $H=S_{11,12}$
\bea
\mathfrak{so}(10,2) & = & \mathfrak{s^{-1}} \oplus \mathfrak{so(10)} + \mathfrak{u(1)}_H  \oplus \mathfrak{s}^{+1}  \nonumber \\
&=& \left( S_{11,i} - i S_{12,i} \right) 
\oplus ( S_{i,j} + S_{11,12} ) \oplus \left( S_{11,i} + i S_{12,i} \right)
\eea
The 10 generators in grade $\pm 1$ subspaces $\mathfrak{s}^{\pm 1} $ can be expressed in terms of the $U(5)$ covariant generators as  follows: 
\bea
\mathfrak{s}^{+1} \Longrightarrow [\left( S_{m,11} + i S_{m,12} \right)= 12 i J^+_{m6} - J^m_6] \quad  \, \oplus \, [ \left( S_{m+5,11} + i S_{m+5,12} \right)= -12 J^+_{m6} + i J^m_6 ] \nonumber \\
\mathfrak{s}^{-1} \Longrightarrow [\left( S_{m,11} - i S_{m,12} \right)= 12 i J_-^{m6} + J^6_m ] \quad \, \oplus \, [ \left( S_{m+5,11} - i S_{m+5,12} \right)= 12 J_-^{m6} + i J^6_m ] \nonumber
\eea
where $m,n,..=1,..5$. They satisfy 
\bea [S_{11,12} , ( S_{11,i} \pm i S_{12,i} ) ] = \pm (S_{11,i} \pm i S_{12,i}  ) \eea
where $i,j=1,\cdots 10 $.
They correspond to the raising and lowering operators of $SO(10,2)$ which we shall label as $S^{\pm}_i$ :
\bea
S^+_i = ( S_{11,i} + i S_{12,i} ) \quad , \quad  S^-_i =( S_{11,i} - i S_{12,i} ) \quad , \quad H = S_{11,12} 
\eea
we have 
\bea 
[H , S^+_i ] =S^+_i \qquad  [ H , S^-_i ] = - S^-_i
\eea
\bea
[ S^-_i , S^+_j  ] =  2 \, \delta_{ij} H + 2 i S_{i,j} 
\eea
\bea
[ S_{i,j}, S^+_k ] = i  \delta_{j,k} \, S^+_i - i \delta_{i,k} \, S^+_j 
\eea
\bea
[ S_{i,j}, S^-_k ] = i  \delta_{j,k} \, S^-_i - i \delta_{i,k} \, S^-_j 
\eea
The raising and lowering operators satisfy
\bea \sum_{i=1}^{10} S^+_i \, S^+_i  =0 \\ 
\sum_{i=1}^{10}  S^-_i \, S^-_i  =0 
\eea
The Casimir operator of $SO(10) $ subgroup is given by 
\bea
C_2(SO(10)) &=& \sum_{i,j=1}^{10} S_{ij} \, S_{ij} \\ 
&= & 2 \sum_{m,n=1}^5 (J^m_n \, J^n_m + 144 J^+_{mn} J_-^{mn}  ) +\frac{10}{9} U^2 -\frac{4}{3} J^6_6 \, U  - 8 J^6_6  + \frac{40}{3} U
\eea

The irreps of $SO(10)$ with the Dynkin labels $ (0,0,0,0,n)$ and $U(1)_H$ charge $(n/2 + 4)$ constructed in appendix B ( see equation \ref{so10irreps})   are annihilated by the action of lowering operators and hence one can generate the basis of a positive energy unitary irreducible representation of $SO(10,2)$ by acting on these states repeatedly with the raising operators $S^+_i$. We shall refer to them simply  as lowest weight irreps of the corresponding unitary irreducible representations of $SO(10,2)$. One can easily determine the $SO(10)\times U(1)_H$ decomposition of the resulting unitary representations of $SO(10,2)$ referred to as the lowest K-types. Since the raising operators $S^+_i$ satisfy
\[ \sum_{i=1}^{10}S^+_i S^+_i =0 \] the products of the raising operators transform as symmetric tensor representations of $SO(10)$ :
\bea 
\underbrace{S^+_i \, S^+_j \, ... S^+_k}_{m} \cong (m,0,0,0,0)_m  
\eea
where the subscript in the Dynkin label of $SO(10)$ refer to the $U(1)_H$ charge. For the lowest  weight vectors of $SO(10)$ given in the previous section we have :
\bea
H  \left(\Tilde{Z}_{78}\right)^n |0\rangle= (\frac{n}{2} +4 ) \left(\Tilde{Z}_{78}\right)^n |0\rangle
\eea
Hence the K-type decomposition of the unitary irreducible representation (UIR) of $SO(10,2)$ with the lowest weight vector \ref{u5lwv} is given by 
\bea
|\Omega_- \rangle &=&  \left(\Tilde{Z}_{78}\right)^n |0\rangle \Longrightarrow \, |(0,0,0,0,n)_{(4+n/2)}\rangle  \, \text{of \,SO(10)} \\ && \Longrightarrow \, \text{ UIR \, of \, SO(10,2) } \equiv  \,  \sum_{m=0}^{\infty} (m,0,0,0,n)_{(m+n/2+4)} \nonumber
\eea

If one interprets $SO(10,2)$ as a conformal group in ten dimensional Minkowskian spacetime and choose 11th coordinate as time coordinate and the space coordinates to run from 2 to 10 we have a noncompact three grading with respect to the dilatation generator $D= S_{1,12}$ such that grade -1 and +1 generators are momentum  $P_\mu$ and special conformal generators $K_\mu$:
\bea
\mathfrak{so}(10,2) = K_\mu \oplus ( S_{\mu,\nu} + D ) \oplus P_\mu 
\eea
where 
\bea 
P_\mu = S_{\mu ,1} -S_{\mu , 12} \\
K_\mu = S_{\mu ,1} +S_{\mu , 12}
\eea
 and $\mu , \nu, ..=2,3,...,11$. They satisfy
 
 \bea 
[ D , P_\mu ] = i P_\mu \qquad  [ D  , K_\mu ]  = - i K_\mu \qquad  
[ P_\mu , K_\nu ] = -2 i (\theta_{\mu\nu} D + S_{\mu \nu} )
\eea
where $S_{\mu \nu} $ are Lorentz group generators in $(9+1)$ dimensions. The Poincare mass operator $P^2=P_\mu P_\nu \theta^{\mu\nu}$  as well as $K^2 = K_\mu K_\nu \theta^{\mu\nu}$ vanish identically as {\it operators in the above realization}:
 \bea
 P^2=P_\mu P_\nu \theta^{\mu\nu} =0 \\
 K^2 = K_\mu K_\nu \theta^{\mu\nu}=0 
 \eea
 Hence the resulting unitary irreps of $SO(10,2)$  describe conformally  massless fields in ten dimensional Minkowskian spacetimes. The relationship between the unitary lowest weight (positive energy)   representations of conformal group and the  conformal fields transforming covariantly under the Lorentz group is established by introducing an intertwiner that interchanges the two pictures \cite{Gunaydin:1999jb,Gunaydin:2017lhg,Chiodaroli:2011pp}. The intertwiner maps the conformal Hamiltonian $H$ into $-D$. Therefore  the UIR of $SO(10,2)$ with lowest weight irrep  with Dynkin labels $(m_1,m_2,m_3,m_4,m_5) $ and $ H=E$ gets intertwined to a conformal field that transforms in the $(m_1,m_2,m_3,m_4,m_5) $ representation under the Lorentz group $SO(9,1)$ with scaling dimension $D=-E$.
 We shall denote the corresponding conformal field as 
 \[ \Phi_D(x_\mu ; (m_1,m_2,m_3,m_4,m_5)) \]
 where $x_\mu$ are the Minkowski coordinates in 10 dimensions.

The lowest weight states  $(\Tilde{Z}_{7,8})^n  |0\rangle$ are eigenstates of the quadratic Casimir of $SO(10,2)$ with eigenvalues
\bea
C_2(SO(10,2))(\Tilde{Z}_{7,8})^n \, |0\rangle = \frac{1}{4} ( n^2 + 6n -16)  \Tilde{Z}_{7,8}^n \, |0\rangle 
\eea
which are also the eigenvalues of the Casimir of the UIRs defined by them,

The corresponding massless conformal fields $  \Phi_{-(4+n/2)}(x_\mu ; (0,0,0,0,n)) $ transform in the representation $(0,0,0,0,n)$ of the ten dimensional Lorentz group $SO(9,1)$. They can be represented as symmetric tensors $\Psi_{\alpha\beta...\gamma}(x_\mu) $  in spinor indices $\alpha , \beta,..=1,2,..,16$ subject to certain constraints under projection to irreps of $SO(9,1)$:
\bea 
 \Phi_{-(4+n/2)}(x_\mu ; (0,0,0,0,n)) \Longleftrightarrow \underbrace{\Psi_{\alpha\beta...\gamma}}_{n }(x_\mu) 
\eea
The field in the $(0,0,0,0,1)$ representation is simply the Weyl spinor $\Psi_\alpha$ in ten dimensions. The field with the Dynkin label $(0,0,0,0,2)$ is a symmetric tensor $\Psi_{\alpha\beta}$ subject to the constraint
\bea
\sigma_\mu^{\alpha\beta} \Psi_{\alpha\beta} =0  \label{constraint} 
\eea
where $\sigma_\mu (\, \mu,\nu=0,1,2,..,9 )$ are the symmetric ten dimensional sigma matrices with coordinate indices $\mu, \nu=0,1,..,9$.  We expect the constraints on higher symmetric tensors to be obtainable  by applying the constraint \ref{constraint} to all possible pairs of spinor indices. 
  
 \section{Distinguished $SU(1,1)_K$ subgroup of $E_{7(-25)}$ and conformal  quantum mechanics}
\label{sec:SU(1,1)_K}

The operator $L_0=\frac{1}{2} (K+M) $ is the compact generator of $SU(1,1)_K$ generated by $ \Delta, M $ and $ K$ and determines its compact three grading 
\bea
{[} L_0 , L_{\pm} {]} = \pm L_{\pm} \\
{[} L_- , L_+ {]} =  2 L_0  
\eea
where $L_-= \frac{1}{2} (-\Delta +M - K ) $ and  $L_+= \frac{1}{2} ( \Delta +M - K ) $

\begin{equation}
L_0  = \frac{1}{4} \left( x^2 + p^2 \right)
     + \frac{1}{2x^2} \mathcal{G}
\end{equation}
where $G= 2(C_{2}(SO(10,2)) + \frac{99}{16})$.  As pointed out in \cite{Gunaydin:2001bt} the eigenvalues of $\mathcal{G}$ determine the coupling constants in the underlying conformal quantum mechanics \cite{deAlfaro:1976je} with the Hamiltonian $L_0$ and the potential
\begin{equation}
V \left( x \right) = \frac{\mathcal{G}}{x^2}
\end{equation}
as well as  the coupling constants in the Hamiltonian of corresponding Calogero models \cite{Calogero:1970nt}. 

Going to the Schrodinger picture the noncompact  generators  of $SU(1,1)_K$  take the form
\begin{equation}
\begin{split}
L_+ &= \frac{1}{4} \left( x - i p \right)^2
   - \frac{1}{2 \, x^2} \mathcal{G}
\\
 &= \frac{1}{4} \left( x - \frac{\partial}{\partial x} \right)^2
    - \frac{1}{2 \, x^2} \mathcal{G}
\\
L_- &= \frac{1}{4} \left( x + i p \right)^2
   - \frac{1}{2 \, x^2} \mathcal{G}
\\
&= \frac{1}{4} \left( x + \frac{\partial}{\partial x} \right)^2
   - \frac{1}{2 \, x^2} \mathcal{G}
\end{split}
\end{equation}
For a given eigenvalue $g$ of $\mathcal{G}$ there exists a wave-function $  \psi_{l_{\alpha_g}}^{\alpha_g} $ in the Hilbert space of the conformal oscillator that is the lowest weight vector of a unitary irreducible representation of $SU(1,1)_K$ :
\bea \psi_{l_{\alpha_g}}^{\alpha_g} \left( x \right)
= C_0 \, x^{\alpha_g} e^{-x^2/2} \eea
where $C_0$ is a normalization constant and
\begin{equation}
\alpha_g
= \frac{1}{2} \pm \sqrt{2 g + \frac{1}{4}} \,.
\end{equation}
One needs to choose the solution for $\alpha_g$ that gives a normalizable wave-function. 
This wave-function satisfies 
\bea
L_-  \psi_{l_{\alpha_g}}^{\alpha_g} \left( x \right) = 0
\\
L_0  \psi_{l_{\alpha_g}}^{\alpha_g} \left( x \right) =  l_{\alpha_g} \psi_{l_{\alpha_g}}^{\alpha_g} \left( x \right)
\eea
where 
\begin{equation}
l_{\alpha_g}
= \frac{\alpha_g}{2} + \frac{1}{4}
= \frac{1}{2} \pm \frac{1}{2} \sqrt{2 g + \frac{1}{4}} \,.
\label{alpha_g}
\end{equation}
 
\textit{We should note that the relationship between the eigenvalue $l_{\alpha_g}$ of $L_0$  and $\alpha_g$ given by the above formula holds only for the lowest weight vector and not for the excited modes of the unitary irrep defined by it. }
By acting on the lowest weight vector $ \psi_{l_{\alpha_g}}^{\alpha_g} \left( x \right) $ of $SU(1,1)_K$ with $L_+$ we create states whose $L_0$ eigenvalues increase in steps of one
\bea
L_0 (L_+)^n \psi_{l_{\alpha_g}}^{\alpha_g} \left( x \right)=(l_{\alpha_g}+n) (L_+)^n  \psi_{l_{\alpha_g}}^{\alpha_g} \left( x \right)
\eea
We shall denote the tensor product of the lowest weight vector $ \psi_{l_{\alpha_g}}^{\alpha_g} \left( x \right) $ of $SU(1,1)_K$ with the Fock vacuum $|0\rangle $ of all the bosonic oscillators as 
\bea
| \psi_{l_{\alpha_g}}^{\alpha_g} \left( x \right);0  \rangle =  \psi_{l_{\alpha_g}}^{\alpha_g} \left( x \right) \otimes |0\rangle \eea

 The values of $g$ are given by the eigenvalues of  $\mathcal{G}= 2(C_{2}(SO(10,2)) + \frac{99}{16}) $. Acting on the state $| \psi_{l_{\alpha_g}}^{\alpha_g} \left( x \right);0  \rangle $  we have \footnote{ This agrees with the general formula for the lowest value of $
g_0 = \frac{1}{8} \left( d - 3 \right) \left( d - 5 \right)
$
 for the minrep of $SO(d,2)$ for. $d>5 \,$ \cite{Fernando:2015tiu}. } 
\bea
C_2(SO(10,2)) | \psi_{l_{\alpha_g}}^{\alpha_g} \left( x \right);0  \rangle  = -4 | \psi_{l_{\alpha_g}}^{\alpha_g} \left( x \right);0  \rangle  \qquad \Longrightarrow  \, \, g =2(-4 +\frac{99}{16})= \frac{35}{8} 
\eea 
Hence 
\bea
\alpha_g = \frac{1}{2} \pm 3
\eea
Normalizability requires the choice $ \alpha_g = \frac{1}{2} +3 =\frac{7}{2} $ and leads to $l_{7/2} = 2$:
\bea
L_0 \,  \psi_{2}^{7/2} \left( x \right) =  2 \, \psi_{2}^{7/2} \left( x \right)
\eea
Hence the tensor product state $ | \psi_{2}^{7/2} \left( x \right);0  \rangle $  is annihilated by all the bosonic annihilation operators and the lowering operator $L_-$ of $SU(1,1)_K$

Consider now states of the form  $\psi_{l_{\alpha_g}}^{\alpha_g}  \Tilde{Z}_{7,8}^n \, |0\rangle $. They are eigenstates of $\mathcal{G}$:
\bea
\mathcal{G} \psi_{l_{\alpha_g}}^{\alpha_g}  \Tilde{Z}_{7,8}^n \, |0\rangle=2 ( C_2 + \frac{99}{16} )  \psi_{l_{\alpha_g}}^{\alpha_g}  \Tilde{Z}_{7,8}^n \, |0\rangle=  \frac{1}{8} \left( 4 n^2 + 24 n + 35 \right) \psi_{l_{\alpha_g}}^{\alpha_g}  \Tilde{Z}_{7,8}^n \, |0\rangle \nonumber
\eea
with eigenvalues $g= \frac{1}{8} \left( 4 n^2 + 24 n + 35 \right)$.
Choosing the positive root of the equation
\bea
 \alpha_g
= \frac{1}{2} \pm \sqrt{2 g + \frac{1}{4}} = \frac{1}{2} \pm (n+3) 
\eea
we get a normalizable state that is the lowest weight vector of $SU(1,1)_K$ 
\bea
L_- \psi_{2+n/2}^{n+7/2}  \Tilde{Z}_{78}^n \, |0\rangle = 0 
\eea
and satisfies
\bea
L_0 \psi_{2+n/2}^{n+7/2} \Tilde{Z}_{78}^n \, |0\rangle = (2+n/2)  \psi_{2+n/2}^{n+7/2} \Tilde{Z}_{78}^n \, |0\rangle \Longrightarrow l_{n+7/2} = 2+n/2
\eea

\section{Compact three-grading of $E_{7(-25)}$ with respect to the $E_6\times U(1)$ subgroup }

Lie algebra of $E_{7(-25)}$ admits a three grading with respect to the Lie algebra of its  maximal compact subgroup $E_6 \times U(1)_R$.

\bea
\mathfrak{e_{7(-25)}} =\mathfrak{g}^{-1}\oplus  \mathfrak{g}^{0}\oplus \mathfrak{g}^{+1}= \mathfrak{27}_{-1} \oplus \mathfrak{e_{6}} + \mathfrak{u(1)}_R \oplus \mathfrak{\bar{27}}_{+1}  \label{comp3grade} 
\eea
The generator of $U(1)_R$ that gives the three-grading is $R$
\bea R = S_{11,12} + \frac{1}{2} ( K + M) = S_{11,12} + L_0 
\eea
and we have
\bea
[ R, \mathfrak{g}^{\pm 1} ]= \pm \mathfrak{g}^{\pm 1}
\eea

The noncompact generators belong to the grade $-1$  and grade $+1$ subspaces and   we shall label as $R_{+}^I$ and $R^{-}_I $ , respectively. 
They consist of the following operators
\bea \mathfrak{g}^{-1} = R^{-}_I =  S^{-}_i \oplus Q^{-}_{\dot{\alpha}} \oplus L_{-} = S^-_i  \oplus  \left((\Tilde{N}_{mn} )  +  (N^{m6}  ) + (N^{78} ) \right) \oplus (L_-)  \label{grade-1} \\
 \mathfrak{g}^{+1} =R_{+}^I=  S^{+}_i \oplus Q_{+}^\alpha \oplus L_{+}=   S^+_i  \oplus \left((N^{mn} )  +  (\Tilde{N}_{m6} ) + (\Tilde{N}_{78} )\right)  \oplus (L_+)  \label{grade+1} 
\eea

 \[ \Tilde{N}_{mn} =1/2 (\Tilde{M}_{mn} - i \Tilde{K}_{mn} ) \quad , \quad  N^{m6}=1/2  (M^{m6} -i K^{m6} ) \] \[ N^{78}=1/2 (M^{78} -i K^{78} ) \quad , \quad L_-= 1/2 (-\Delta +M - K) \] 
\[ N^{mn}=1/2 (M^{mn} + iK^{mn} )  \quad ,\quad \Tilde{N}_{m6}=1/2 (\Tilde{M}_{m6} +i \Tilde{K}_{m6} ) \]
 \[\Tilde{N}_{78}=1/2 (\Tilde{M}_{78} +i \Tilde{K}_{78} ) \quad , \quad L_+=1/2(\Delta +M -K)  \]
where $I,J,..=1,2,..,27$ , $\alpha , \dot{\alpha} =1,2 \cdots , 16. $ ,   $i,j,..=1,..10$ and $m,n,..=1,..5$. 
The generators in grade $\pm 1$ subspaces decompose under the $SO(10)$ subgroup as follows
\bea
\mathfrak{g}^{-1} \quad \Longrightarrow  \quad \bar{27}= 10^{-2}+ \bar{16}^1 + 1^4 \\
\mathfrak{g}^{+1} \quad \Longrightarrow  \quad 27 = 10^2 + 16^{-1} + 1^{-4} 
\eea
where the superscripts denote the charges with respect to the operator $3F$ where $F=2/3 (S_{(11,12)}-2 L_0)$ is the generator that determines the 3-grading of compact $E_6$ given in Appendix C. 
Hence we have the following identification;
\bea
 R_{+}^i =S^+_i , \quad i=1,..,10 ; \quad R_{+}^{10+\alpha}=Q_{+}^\alpha , \quad \alpha=1,2,..,16 , ; \quad R_{+}^{27} =L_+ \\
  R^{-}_i =S^-_i , \quad i=1,..,10 ; \quad R^{-}_{10+\dot{\alpha}}=Q^{-}_{\dot{\alpha}} , \quad \dot{\alpha} =1,2,..,16 ; \quad R^{-}_{27} =L_-
 \eea
 with $\alpha$ and $\dot{\alpha}$  denoting the spinor indices of $SO(10)$ and its conjugate, respectively.
Since $S^-_i$ and $Q^{-}_{\dot{\alpha}}$ annihilate the Fock vacuum $|0\rangle$ and the state $\psi_{2}^{7/2} \left( x \right)$ is annihilated by $L_-$ we see that the tensor product state $ | \psi_{2}^{7/2} \left( x \right);0  \rangle $ is the lowest weight vector of the minimal unitary representation of $E_{7(-25)}$. It satisfies 
\bea
R | \psi_{2}^{7/2} \left( x \right);0  \rangle &=& (L_0 + S_{11,12} )| \psi_{2}^{7/2} \left( x \right);0  \rangle = 6 | \psi_{2}^{7/2} \left( x \right);0  \rangle \\
R^{-}_I | \psi_{2}^{7/2} \left( x \right);0  \rangle  &=&0  \quad ,  I=1,2,\cdots , 27 
\eea 
The lowest weight vector  $ | \psi_{2}^{7/2} \left( x \right);0  \rangle $ is a singlet of the compact subgroup $E_6$ whose commutation relations and Casimir are given in the appendix C.

 \section{Decomposition of the minrep of  $E_{7(-25)} $ with respect to the $SO(10,2)\times SU(1,1)_K$ subgroup and massless conformal fields in ten dimensional Minkowskian space-time } 
 
 The Hilbert  space of the minimal unitary representation of $E_{7(-25)}$ is spanned by states belonging to  the tensor product of the Fock space of the bosonic oscillators and the Hilbert spaces of a family of  conformal quantum mechanics whose coupling constants are determined by the eigenvalues of the Casimir operator of $SO(10,2)$ subgroup. 
 The lowest weight vector   $ | \psi_{2}^{7/2} \left( x \right);0  \rangle $ of the minrep of $E_{7(-25)}$. satisfies  
\bea
\Tilde{Z}_{mn}  | \psi_{2}^{7/2} \left( x \right);0  \rangle  =0 \qquad  \, Z^{m6}  | \psi_{2}^{7/2} \left( x \right);0  \rangle  =0  \qquad  Z^{78}  | \psi_{2}^{7/2} \left( x \right);0  \rangle  =0 \\
L_{-}  | \psi_{2}^{7/2} \left( x \right);0  \rangle  =1/2(-\Delta +M - K)  | \psi_{2}^{7/2} \left( x \right);0  \rangle =0 \nonumber
\eea
and is a singlet of the maximal compact subgroup  $E_6$  and has the $U(1)_R$ charge 6:
\bea
R  | \psi_{2}^{7/2} \left( x \right);0  \rangle  = 6  | \psi_{2}^{7/2} \left( x \right);0  \rangle 
\eea

Acting on the lowest weight vector $ | \psi_{2}^{7/2} \left( x \right);0  \rangle $ with the 27 grade +1 operators repeatedly  we generate an infinite number of states that form the basis of the minrep of $ E_{7(-25)}$ :
\bea
\left(
| \psi_{2}^{7/2} \left( x \right);0  \rangle \oplus R_+^I | \psi_{2}^{7/2} \left( x \right);0  \rangle  \oplus R_+^I R_+^J | \psi_{2}^{7/2} \left( x \right);0  \rangle \oplus \cdots  \right) \,  \Leftrightarrow \mathrm{Minrep \, \, of } \, E_{7(-25)}
\eea
In Dynkin labels the K-type decomposition of the minrep of $E_{7(-25)}$ with respect to its maximal compact subgroup $E_6\times U(1)_R$ is given by
\bea
\mathrm{Minrep \, \, of } \, E_{7(-25)} \equiv \sum_{n=0}^\infty \bigoplus ( n,0,0,0,0,0)_{(n+6)}  
\eea
where the subscript indicates the $U(1) $ $R$-charge. 
First few levels the irreps of $E_6$ have dimensions
\bea
(1,0,0,0,0,0) = 27  \\
(2,0,0,0,0,0) = \overline{351}'  \\
(3,0,0,0,0,0) = 3003  \\
(4,0,0,0,0,0) = 19305' 
\eea
using the conventions of \cite{Feger:2019tvk}.

 We shall study the decomposition of  the minrep of  $E_{7(-25)} $ with respect to its $SO(10,2) \times SU(1,1)_K$ subgroup and show that it decomposes into infinitely many irreps belonging to the holomorphic discrete ( lowest weight) representations of $SO(10,2) \times  SU(1,1)_K$ .
Decomposing the grade $+1$  generators $R_{+}^I$ and grade $-1$ generators as $R^{-}_I $ with respect to its $SO(10)$ subgroup we have 
\bea
R_{+}^I = S^{+}_i \oplus Q_{+}^\alpha \oplus L_{+} =10 + 16 +1 \\
R^{-}_I = S^{-}_i \oplus Q^{-}_{\dot{\alpha}} \oplus L_{-} = 10 +\bar{16} + 1 \
\eea
where 
\bea
Q_+^\alpha \leftrightarrow ( (M^{mn} + iK^{mn} ) + (\Tilde{M}_{m6} +i \Tilde{K}_{m6} ) + (\Tilde{M}_{78} +i \Tilde{K}_{78} ) ) \\
Q^-_{\dot{\alpha}}  \leftrightarrow ((\Tilde{M}_{mn} - i \Tilde{K}_{mn} ) +  (M^{m6} -i K^{m6} ) + (M^{78} -i K^{78} ) )
\eea
with the indices $\alpha, \beta, \cdots $ running from 1 to 16 corresponding to the Weyl spinor representation $16$  of $SO(10)$ and the dotted indices correspond to the conjugate spinor indices. The generators $L_+$ and $L_-$  of $SU(1,1)_K$ 
and generators $S^{+}_i$ and $S^-_i$  of $SO(10,2)$ 
 carry the following charges with respect to $H$ and $L_0$:
\bea
[H, S^{\pm}_i ] = \pm S^{\pm}_i \quad , \quad  [L_0, S^{\pm}]=0 \\ \nonumber
[H, Q_+^\alpha ] =1/2 \, Q_+^\alpha \quad , \quad [L_0,Q_+^\alpha ] =1/2 \,  Q_+^\alpha \nonumber \\ \nonumber
[H,Q^-_{\dot{\alpha}}  ] = -1/2 \, Q^-_{\dot{\alpha}}  \quad , \quad [L_0, Q^-_{\dot{\alpha}}  ]=-1/2 \, Q^-_{\dot{\alpha}}  \\
{[}H , L_{\pm} {]} =0 \quad , \quad {[} L_0, L_{\pm} {]} = \pm L_{\pm}
\eea
The lowest weight vector $ | \psi_{2}^{7/2} \left( x \right);0  \rangle $ of the minrep of $E_{7(-25)}$  is simultaneously the lowest weight vector of  an irrep of $SU(1,1)_K$ and  an irrep of $SO(10,2)$. It satisfies
\bea
L_0 | \psi_{2}^{7/2} \left( x \right);0  \rangle = 2 | \psi_{2}^{7/2} \left( x \right);0  \rangle \\
H | \psi_{2}^{7/2} \left( x \right);0  \rangle = 4 | \psi_{2}^{7/2} \left( x \right);0  \rangle \\
S_{ij}| \psi_{2}^{7/2} \left( x \right);0  \rangle = 0  \quad , \quad i,j=1,..,10 \\
L_- | \psi_{2}^{7/2} \left( x \right);0  \rangle =0 
\eea
We shall denote the  states that transform in the representation of $SO(10)$ with Dynkin labels $(m_1,m_2,m_3,m_4,m_5)$ and $U(1)_H$ charge $E$  as $| \Psi_E (m_1,m_2,m_3,m_4,m_5)  \rangle $ 
and its tensor product  with  conformal wave function $\psi^{\alpha_g}_{l_{\alpha_g}}(x) $ with $U(1)_{L_0}$ charge $l_{\alpha_g}$ as
\bea
 | \psi^{\alpha_g}_{l_{\alpha_g}} (x);\Psi_E (m_1,m_2,m_3,m_4,m_5)  \rangle
 \eea
In this notation the lowest weight vector $ | \psi_{2}^{7/2} \left( x \right);0  \rangle $  takes the form
\[  | \psi^{7/2}_2(x); \Psi_4 ((0,0,0,0,0) ) \rangle \]
Acting on the state $| \psi_{2}^{7/2} \left( x \right);0  \rangle $ with the raising operators $S^+_i$ repeatedly one generates the basis of the minimal unitary representation of $SO(10,2)$ that describes a massless scalar field in ten dimensional Minkowskian spacetime. Its K-type decomposition with respect to its maximal compact subgroup $SO(10) \times U(1)_H$ is as follows
\bea
\sum_{n=0}^{\infty} S_{i_1}^+ S_{i_2}^+ \cdots S_{i_n}^+ | \Psi_4 ((0,0,0,0,0) ) \rangle = \sum_{n=0}^\infty \bigoplus  | \Psi_{4+n}  ((n,0,0,0,0) ) \rangle
\eea
Acting with the raising operators $L_+$ and $S^+_i$  repeatedly on the lowest weight vector  $ | \psi^{7/2}_2(x); \Psi_4 ((0,0,0,0,0) ) \rangle$ one obtains  the higher modes of the unitary irrep of  $SO(10,2)\times SU(1,1)_K$ : 
\bea
(L_+) ^m S_{i_1}^+ S_{i_2}^+ \cdots S_{i_p}^+ | \psi^{7/2}_2(x); \Psi_4 ((0,0,0,0,0) ) \rangle = | \psi^{7/2}_{m+2}(x); \Psi_4 ((p,0,0,0,0) ) \rangle
\eea

On the other hand by acting on the lowest weight vector  $ | \psi^{7/2}_2(x); 0) \rangle$ with the generators $Q_+^\alpha$ we generate a new lowest weight irrep  of $SO(10,2)\times SU(1,1)_K $ 
\bea
S^-_i Q_\alpha^+ | \psi^{7/2}_2(x); 0 \rangle =0  \quad , \quad L_- Q_\alpha^+ | \psi^{7/2}_2(x); 0 \rangle =0 
\eea
that transform in the spinor representation $16=(0,0,0,0,1)$ of $SO(10)$  with $E=9/2$ and has $U(1)_{L_0}$ charge $l_{\alpha_g}=5/2$. The true lowest weight vector of this lowest weight irrep is simply the state
\bea
\Tilde{N}_{78} | \psi^{7/2}_2(x); 0 \rangle = 1/2((x - i p)+7/(2x) ) \Tilde{Z}_{78} | \psi^{7/2}_2(x); 0 \rangle= \Tilde{Z}_{78} | \psi^{9/2}_{5/2}(x); 0 \rangle
\eea
where we used the identity
\bea
 1/2((x - i p)+\frac{\alpha}{x} ) \psi^{\alpha}_{l_{\alpha_g}}(x) = \psi^{\alpha+1}_{(l_{\alpha_g}+1)} (x)
 \eea
 Note that 
\bea
 L_- \psi^{\alpha+1}_{l_{\alpha+1}} (x) =0 \quad , \quad L_0 \psi^{\alpha+1}_{l_{\alpha +1}} (x) = (\frac{(\alpha +1)}{2} + \frac{1}{4} )\psi^{\alpha+1}_{l_{\alpha +1}}(x)  
 \eea
and as we saw in section 
\ref{covariant}  the state $\Tilde{Z}_{78} |0\rangle$ is the lowest weight vector of the spinor representation $16$ of $SO(10)$ in the bosonic Fock space. 
By acting with higher powers of $\Tilde{N}_{78}$ on the lowest weight vector $|\psi^{7/2}_2(x); 0 \rangle $ we generate new lowest weight irreps of $SO(10,2)$:
\bea
(\Tilde{N}_{78} )^p | \psi^{7/2}_2(x); 0 \rangle = (\Tilde{Z}_{78})^p | \psi^{\alpha=7/2+p}_{l_\alpha=2+p/2} (x);0\rangle 
\eea
The corresponding unitary irreps  describe infinitely many conformally massless representations of the conformal group $SO(10,2)$ in ten dimensional Minkowskian spacetime.

The irreps $(n,0,0,0,0,0)_{(n+6)}$ of the maximal compact subgroup $E_6\times U(1)_R$ decomposes under its subgroup $SO(10)\times U(1)_H \times U(1)_{L_0}$ as follows:
\bea
&E_6\times U(1)_R \subset SO(10)\times U(1)_H \times U(1)_{L_0}: \\
&(n,0,0,0,0,0)_{n+6} = \sum_{(m+p+q)= n}  (m,0,0,0,p)\times (H=4+m+p/2) \times (L_0=p/2 + q +2) \nonumber \label{so10decomp}
\eea
Let us list the decomposition of the first four levels
\bea
(1,0,0,0,0,0)_{7} &=& (0,0,0,0,0)(4)(3) + (1,0,0,0,0)(5)(2) + (0,0,0,0,1)(9/2)(5/2)  \\
(2,0,0,0,0,0)_{8}&=& (0,0,0,0,0)(4)(4) + (1,0,0,0,0)(5)(3) + (0,0,0,0,1)(9/2)(7/2)  \nonumber \\  && {\bf  + (2,0,0,0,0)(6)(2)+ (0,0,0,0,2)(5)(3) + (1,0,0,0,1)(11/2)(5/2)} \nonumber \\ \nonumber
(3,0,0,0,0,0)_{9} &=& (0,0,0,0,0)(4)(5) + (1,0,0,0,0)(5)(4) + (0,0,0,0,1)(9/2)(9/2) \nonumber
  \\&& + (2,0,0,0,0)(6)(3) + (0,0,0,0,2)(5)(4) + (1,0,0,0,1)(11/2)(7/2)\nonumber \\&& {\bf  + (3,0,0,0,0)(7)(2)   + (0,0,0,0,3)(11/2)(7/2) + (2,0,0,0,1)(13/2)(5/2) }\nonumber \\&& {\bf  + (1,0,0,0,2)(6)(3)}  \nonumber  \\
  (4,0,0,0,0,0)_{10} &=&(0, 0, 0, 0, 0)(4)(6) +(1, 0, 0, 0, 0)(5)(5) + (0, 0, 0, 0, 1)(9/2)(11/2) \nonumber \\
  && (2, 0, 0, 0, 0)(6)(4) + (0, 0, 0, 0, 2)(5)(5)+(1, 0, 0, 0, 1)(11/2)(9/2) \nonumber \\ &&
  (3, 0, 0, 0, 0)(7)(3) +(0, 0, 0, 0, 3)(5)(5) +  (2, 0, 0, 0, 1)(11/2)(9/2) \nonumber \\ &&
  + (1, 0, 0, 0, 2)(6)(4)  \nonumber  \\ && {\bf + (4, 0, 0, 0, 0)(8)(2) +
  (0, 0, 0, 0, 4)(6)(4) + (2, 0, 0, 0, 2)(7)(3) }  \nonumber \\ && {\bf +(1, 0, 0, 0, 3)(13/2)(7/2) + (3, 0, 0, 0, 1)(15/2)(5/2)}\nonumber
  \eea
 where the representations that are not simply $SU(1,1)_K$  excitations of the representations that occur in the previous level are indicated boldface.
 
  In the decomposition of the minrep of $E_{7(-25)}$ given in \ref{so10decomp} only the states with the $SO(10)\times U(1)_H \times U_{L_0}$ labels of the form
  \[ (0,0,0,0,p)\times(H=4+p/2)\times (L_0=2+p/2) \] 
  describe simultaneous lowest weight irreps of $SO(10,2)$ and $SU(1,1)_K $ subgroups. 
  They correspond to the tensor product states 

  \bea
 | \psi^{\alpha=7/2+p}_{l_\alpha=2+p/2} (x);\Psi_{(4+p/2)} (0,0,0,0,p)  \rangle
 \eea

   If we denote the Hilbert spaces of the UIRs of $SU(1,1)_K$ and $SO(10,2)$ with lowest weight irreps $
  \psi^{\alpha=7/2+p}_{l_\alpha=2+p/2} (x)$ and $|\Psi_{(4+p/2)} (0,0,0,0,p)  \rangle$ as $\mathcal{H}(\psi^{\alpha=7/2+p}_{l_\alpha=2+p/2} (x)$ and $\mathcal{H}(\Psi_{(4+p/2)} (0,0,0,0,p)  \rangle)$, respectively, then we can write down the decomposition of the Hilbert space of the minrep of  $E_{7(-25)}$ as
 \bea
  \mathcal{H}_{minrep}(E_{7(-25)})= \sum_{p=0}^\infty  \bigoplus \left( \mathcal{H}(\Psi_{(4+p/2)} (0,0,0,0,p) \bigotimes \mathcal{H}(\psi^{\alpha=7/2+p}_{l_\alpha=2+p/2} (x) ) \right)
  \eea
  
  The results above  show that the spectrum of the minimal unitary representation of $E_{7(-25)}$ decomposes into infinitely many irreps of $SO(10,2)\times SU(1,1)_K$ and one can interpret the minrep as an infinite multiplet with {\it bosonic supersymmetry}  generators  transforming in spinor representation $16$ of $SO(10)$ and its conjugate. Each irrep of $SO(10,2)$ inside the minrep of $E_{7(-25)}$ comes with infinite multiplicity which form an irrep of the " bosonic R-symmetry group" $SU(1,1)_K$. This provides an example of bosonic symmetry algebra whose unitary spectrum exhibits spacetime supersymmetry\footnote{ The possibility that a  bosonic symmetry algebra may exhibit spacetime supersymmetry in its spectrum was first mentioned to me by Edward Witten in connection with hyperbolic Lie algebra $E_{10}$ (Cambridge, 1985). Whether this can be realized within the framework of attempts to relate the hyperbolic Lie algebra $E_{10}$ to M-theory is still an open problem\cite{KN24}. For recent reviews  on the subject and further references see \cite{ Kleinschmidt:2022qwl,Nicolai:2024hqh}. }. 
  
  The results obtained for $E_{7(-25)}$ can be truncated to noncompact groups  $SO^*(12),SU(3,3)$ and $Sp(6,\mathbb{R})$ which are the conformal groups of  simple Euclidean Jordan algebras of  degree three over the quaternions, complex numbers and real numbers. These truncations are discussed in appendix D.    
  

\section{ Deformations of the minimal unitary representation of $E_{7(-25)}$ and massless conformal fields in ten dimensions}
In this section we shall study deformations of the minrep of $E_{7(-25)}$ along the lines of deformations of the minreps of conformal groups  $SO(d+2,2)$ studied in \cite{Fernando:2015tiu}. The $SO(10,2)$ subgroup of $E_{7(-25)}$ admits a three-grading with respect to its maximal rank subgroup $SU(5,1)\times U(1)$ and we shall realize its generators as bilinears of two sets of bosonic oscillators transforming in the fundamental representation of $U(5,1)$ 
\bea
{[} X_a, P^b  {]} = \delta_a^b \quad  , \quad  {[} Q^a , Y_b  {]} =\delta^a_b 
\eea
subject to the Hermitian conjugation properties
\bea
X_a^\dagger = \eta_{ab} P^b \quad , \quad (Q^a)^\dagger = \eta^{ab} Y_b 
\eea 
where $a,b=1,2,...,6$ and $\eta_{ab} = \mathrm{ Diag} (1,1,1,1,1,-1) $. Following bilinears generate $SU(5,1) $ under commutation:
\bea 
L^a_b = P^a X_b -Q^a Y_b -\frac{1}{6} \delta^a_b  (P^c X_c - Q^c Y_c )  
\eea
 The additional generators of $SO(10,2)$  are given by
\bea
L_+^{ab} = P^a Q^b - P^b Q^a \quad , \quad L^-_{ab} = X_a Y_b - X_b Y_a 
\eea
They satisfy 
\bea
{[} L^-_{ab} , L_+^{cd} {]} =  \delta^a_b L^d_a + \delta^d_a L^c_b-  \delta^c_a L^d_b - \delta^d_b L^c_a - \frac{1}{3} (\delta^c_a \delta^d_b -\delta^d_a \delta^c_b)  \hat{L} 
\eea
where $\hat{L}= \frac{1}{2} ( P^a X_a + X_a P^a - Q^a Y_a - Y_a Q^a ) $. 
The quadratic Casimir of $SO(10,2)$ generated by them is
\bea \label{CasO}
C_2(O) = \frac{1}{6} L^a_b L^b_a + \frac{1}{12} ( L^-_{ab} L_+^{ab} + L_+^{ab} L^-_{ab} ) + \frac{1}{36} \hat{L}^2
\eea
We shall label the $SO(10,2)$ generated by the above operators as $SO(10,2)_O$. 

We can extend the generators of $SO(10,2) $ given in section \ref{minrep-e7-25} with the above generators of $SO(10,2)_O$ :
\bea
T^-_{ab} = J^-_{ab} + \frac{1}{12} L^-_{ab}  \quad , \quad T_+^{ab} = J_+^{ab} + \frac{1}{12} L_+^{ab}  \quad , \quad T^a_b = J^a_b + L^a_b 
\eea
They satisfy the commutation relations of $SO(10,2)$: 
\bea
[ T^-_{ab} , T_+^{cd} ] &=& \frac{1}{144}   ( \delta^c_b T^d_a - \delta^c_a T^d_b -\delta^d_b T^c_a + \delta^d_a T^c_b)- \frac{1}{108} \delta^{cd}_{ab} ( U+\frac{1}{2} \hat{L} ) \\
{[} U+\frac{1}{2} \hat{L}, T^a_b {]} &=& 0 \quad , \quad {[} U+\frac{1}{2} \hat{L}, T^-_{ab} {]} = -  T^-_{ab} \quad , \quad  {[}U+\frac{1}{2} \hat{L}, T^{ab}_+ {]}=  T^{ab}_+
 \\
{[} T^-_{ab} , T^-_{cd} {]} &=& {[} T_+^{ab} , T_+^{cd} {]} =0
\eea
We shall label the Lie group generated by $T^-_{ab}, T_+^{ab}, U$ and $T^a_b$ as $SO(10,2)_T $. 
The quadratic Casimir of  $SO(10,2)_T$ is given by
\bea
C_2(T) = \frac{1}{6} T^a_b T^b_a +12 ( T^-_{ab} T_+^{ab} + T_+^{ab} T^-_{ab} ) + \frac{1}{9} (U + \frac{1}{2}\hat{L} )^2
\eea
To deform the minrep of $E_{7(-25)}$ we shall replace the generators of $SO(10,2)$ subgroup with those of $SO(10,2)_T$ 
while keeping  the grade -2 and -1 generators in the  quasiconformal realization   unchanged.  We  have 
\bea
[ T^a_b , M^{cd} ] = \delta^c_b M^{ad} + \delta^d_b M^{ca} -\frac{1}{3} \delta^a_b M^{cd}  \\
{[} T^a_b , \Tilde{M}_{cd} {]} = - \delta^a_c \Tilde{M}_{bd} - \delta^a_d \Tilde{M}_{cb} +\frac{1}{3} \delta^a_b \Tilde{M}_{cd} \\
{[} T^a_b , M^{78} {]} =0 \quad , \quad {[}T^a_b , \Tilde{M}_{78} {]} =0 
\eea
where $a,b,..=1,..,6$. 
 On the other hand the grade $+2 $ and $+1$ generators get modified together with grade zero generators. We shall denote the deformed generators of grade +2  and +1 subspaces of the Lie algebra $\mathfrak{e}_{7(-25)}$ with a prime. We have
\bea
K'  = \frac{1}{2} p^2 + \frac{2}{x^2} \left( 3 C_2(T) - 2 C_2 -\frac{3}{2} C_2(O) +\frac{99}{16} \right) 
\eea
\bea
K'^{ab} = -p Z^{ab} + \frac{2i}{x} \left(  {[} Z^{ab}, ( 3 C_2(T) -2 C_2 -3/2  C_2(O) {]}  \right) \\
\Tilde{K'}_{ab} = - p \Tilde{Z}_{ab} + \frac{2i}{x} \left( {[} \Tilde{Z}_{ab} ,( 3 C_2(T) - 2 C_2 -3/2 C_2(O) {]} \right) \\
K'^{78} = -p Z^{78} + \frac{2i}{x} \left(  {[} Z^{78}, ( 3 C_2(T) -2 C_2 -3/2  C_2(O) {]}  \right) \\
\Tilde{K'}_{78} = - p \Tilde{Z}_{78} + \frac{2i}{x} \left( {[} \Tilde{Z}_{78} ,( 3 C_2(T) - 2 C_2 -3/2 C_2(O) {]} \right)
\eea
They satisfy
\bea
[\Tilde{M}_{ab} , K' ] = -i \Tilde{K'}_{ab} \quad , \quad [ \Tilde{K'}_{ab} , M ] =i \Tilde{M}_{ab}  \\
{[} K'^{ab} , M {]}=  i M^{ab}  \quad , \quad  {[} M^{ab}, K' {]} = - i K'^{ab} 
\eea
\bea
[ M^{ab} , K'^{cd} ] = - 6 i \epsilon^{abcdef} T^-_{ef} 
\eea
\bea
[ M, K' ] = - \Delta  \quad , \quad [ \Delta , K' ] =  2 K' \quad , \quad [ \Delta , M ] = - 2 M 
\eea
\bea
[ \Tilde{M}_{ab} , K'^{cd} ] = -i \left( \delta^c_a T^d_b + \delta^d_b T^c_a - \delta^d_a T^c_b - \delta^c_b T^d_a \right)
- i \delta^{cd}_{ab} \left( \Delta -\frac{2}{3} ( U + \frac{1}{2} \hat{L} ) \right) 
\eea
\bea 
[ M^{78} , K'^{cd} ] = - 12 i T_+^{cd} \quad , \quad [ M^{cd} , K'^{78} ] = -12 i T_+^{cd} 
\eea
The  quadratic Casimir of the deformed  minimal unitary realization of
$\mathfrak{e}_{7(-25)}$  is no longer a c-number and is given by 
\begin{equation}
\begin{split}
  \mathcal{C'}_2 \left( \mathfrak{e}_{7(-25)} \right) &= \mathcal{C'}_2 \left( \mathfrak{so}(10,2)_T \right) +
               \frac{1}{12} \Delta^2 + \frac{1}{6} \left( M K' + K' M \right) \\
                & - \frac{i}{12} \left( \Tilde{M}_{ab} K'^{ab} +  K'^{ab} \Tilde{M}_{ab}
                                          - \Tilde{K'}_{ab}  M^{ab} - M^{ab} \Tilde{K'}_{ab} \right)   \\
               & - \frac{i}{6} \left( \Tilde{M}_{78} K'^{78} +  K'^{78} \Tilde{M}_{78}
                                          - \Tilde{K'}_{78} M^{78} - M^{78} \Tilde{K'}_{78} \right) \\
               & = \frac{3}{2} C_2(O)-14 
\end{split}
\end{equation}
where $C_2(O)$ is the Casimir of $SO(10,2)$ generated by the deforming oscillators $(X,P)$ and $(Y,Q)$  given in equation \ref{CasO}.  
Fock space of these oscillators decompose into infinitely many unitary irreps of $SO(10,2)_T$ and each irrep describes  a deformation of the minrep of $E_{7(-25)}$. These representations of $SO(10,2)_T$ correspond, in general, to massive representations as ten dimensional conformal group. However when the generators of $SO(10,2)_O$ are restricted  to the  $SO(8)$ subgroup one finds that the Poincare mass operator $P\mu P^\mu$ vanishes identically. On the other hand we know that the minrep of $SO(10,2)$ admits an infinite family of deformations describing massless conformal fields in ten dimensions labelled by the Casimir of its  $SO(8)$ subgroup in its quasiconformal realization using its 5-graded decomposition with respect to $SO(8)\times SU(1,1)$ subgroup\cite{Fernando:2015tiu}.  This suggests that vanishing of the Poincare mass when restricted to the $SO(8)$ subgroup above may be related to deformations of the minrep of $SO(10,2)$ subgroup. This question and   a detailed study of the deformations of the minrep of $E_{7(-25)}$ and their decompositions with respect to the $SO(10,2)_T\times SU(1,1)_K$ subgroup will be the subject of a separate study.  

\section{Decomposition of the minimal unitary representation  of $E_{7(-25)}$ with respect to its $SO^*(12)\times SU(2)$ subgroup.  }

 The group $E_{7(-25)}$ has another maximal subgroup, namely $SO^*(12)\times SU(2)$.  As stated above the Lie algebra  of $SO^*(12)$ is the conformal group of the spacetime coordinatized by $J_3^{\mathbb{H}}$ and its Lorentz group is $SU^*(6)$\cite{Gunaydin:1992zh,Gunaydin:2005zz}. It admits a compact three-grading with respect to the Lie algebra  $U(6)$ of its maximal compact subgroup which we denote as:
 \eq
 \mathfrak{so}^*(12) =  \mathfrak{h}^{-1}  \oplus  \mathfrak{h}^{0} \oplus 
   \mathfrak{h}^{+1}
 \en
where $\mathfrak{h}^{0}=\mathfrak{su}(6)+\mathfrak{u}(1) $. The generators of $U(6)$ subgroup are given by
\bea
H^m_n &=&J^m_n \quad , m,n=1,..., 5 \\
H^6_6 &=&-\frac{1}{2}(M+K) -J^6_6 +\frac{1}{3} U \\
H^6_m &=&\sqrt{2} \,  C^{m6}=  \frac{1}{\sqrt{2}} ( M^{m6} + i K^{m6}) \quad,  m=1,..., 5 \\
H^m_6 &=& \sqrt{2} \, \Tilde{C}_{m6} =-\frac{1}{\sqrt{2}} ( \Tilde{M}_{m6} - i \Tilde{K}_{m6}) \quad , m=1,..., 5 
\eea
We shall denote the $SU(6)$ generators as $W^a_b$  defined by
\bea
W^a_b = H^a_b -\frac{1}{6} \delta^a_b  \sum_{c=1}^6 H^c_c 
\eea 
They satisfy the commutation relations
\bea [ W^a_b , W^c_d ] = \delta_b^c W^a_d - \delta^a_d W^c_b 
\eea
The generators belonging to the grade -1  subspace are as follows
\bea
H^-_{mn}=\sqrt{2} \, \Tilde{N}_{mn} =\frac{1}{\sqrt{2}} (\Tilde{M}_{mn} - i \Tilde{K}_{mn}) \quad ,  \quad m,n=1,...,5 \\
H^-_{m6}= J^6_m \quad , \quad H^-_{6m}=-J^6_m  \quad \quad  \nonumber
\eea
and those in grade +1  subspace are
\bea
H_+^{mn}=\sqrt{2} \, N^{mn}  = \frac{1}{\sqrt{2}} ( M^{mn} +i K^{mn} ) \quad ,  \quad m,n=1,..,5 \\
H_+^{m6}= - J^m_6 \quad ,H_+^{6m}= J^m_6 , \quad m=1,..,5  \\
\eea
They satisfy the commutation relations
\bea
{[} H_+^{ab},H^-_{cd} {]} &=& \delta^a_d W^b_c +\delta^b_c W^a_d - \delta^a_c W^b_d -\delta^b_d W^a_c  - \frac{4}{3} \delta^{ab}_{cd} R  \\
{[} W^a_b , H_+^{cd}{]} &=& \delta^c_b H_+^{ad} + \delta^d_b H_+^{ca} -\frac{1}{3} \delta^a_b H_+^{cd} \\
{[} W^a_b , H_{cd}^-{]} &=& - \delta^a_c H_{bd}^-  - \delta^a_d H_{cb}^- +\frac{1}{3} \delta^a_b H_{cd}^- 
\eea
where \[R = \frac{1}{2} ( M+K -\frac{2}{3} U - 2 J^6_6 )\] and $\delta^{ab}_{cd} = \frac{1}{2} (\delta^a_c \delta^b_d - \delta^a_d \delta^b_c)$ with $a,b,..=1,2,3,4,5,6$.
Note that $R $ can be written as
\bea
R =\frac{1}{2} ( M+K -\frac{2}{3} U - 2 J^6_6 )=\frac{1}{2} ( M+K+ZN +8) = L_0 + S_{11,12} 
\eea
where $ZN$ is the number operator for all the bosonic oscillators
\[ ZN := Z^{m,n} \Tilde{Z}_{m,n} -2 \Tilde{Z}_{m,6} Z^{m,6} - 2 \Tilde{Z}_{7,8} Z^{7,8} \]
The generator $R $ determines the three grading of $SO^*(12)$ as well 
\bea
{[} R  , W^a_b {]}=0 , \quad {[}R ,H_+^{ab}{]} = H_+^{ab} , \quad {[} R , H^-_{ab}{]} = - H^-_{ab} 
\eea 
where $ a,b,..=1,...,6$.  The quadratic Casimir of $SO^*(12)$ subgroup is given by
\bea
C_2(SO^*(12))&=& W^a_b W^b_a +\frac{2}{3} (R )^2 -\frac{1}{2} ( H_+^{ab}\, H^-_{ab} +  H^-_{ab} \, H_+^{ab} ) \\
&=& W^a_b W^b_a +\frac{2}{3} (R )^2  - H_+^{ab}\, H^-_{ab}  -10 R  
\eea
The generators of $SU(2)$ subgroup that centralizes $SO^*(12)$ are given by
\bea
J^+ =\sqrt{2} \, C^{78} = \frac{1}{\sqrt{2}}(M^{78}  + i K^{78}) \\
J^-= \sqrt{2} \, \Tilde{C}_{78}  =- \frac{1}{\sqrt{2}} (\Tilde{M}^{78} -i \Tilde{K}^{78})\\
J^3 = \frac{1}{4} (M+K) +\frac{1}{2} U 
\eea
They satisfy the commutation relations
\bea
{[}J^+,J^-{]}= 2 J^3 \\
{[}J^3, J^+{]} = J^+ \\
{[}J^3, J^-{]} = - J^- 
\eea

The generators belonging to the coset $E_{7(-25)}/SO^*(12)\times SU(2)$  decompose as follows with respect to the $SO^*(12)\times SU(2)$ subgroup
\bea
E_{7(-25)}/SO^*(12)\times SU(2) \Longrightarrow  (32,2)
\eea
and further with respect to the  $U(6)\times SU(2)$ subgroup as
\bea
E_{7(-25)}/SO^*(12)\times SU(2) \Longrightarrow    (6,2)_{nc} \oplus (20,2)_{c} \oplus (\bar{6},2)_{nc}  
\eea
where $(6,2)_{nc} \oplus (\bar{6},2)_{nc} $ are noncompact generators belonging to the coset $SU(6,2)/U(6)\times SU(2))$ and the compact generators $(20,2)_{c}$ belong to the compact coset $E_6/SU(6)\times SU(2)$. The noncompact generators in $27 $ and $\bar{27}$ of compact subgroup $E_6$ decompose as follows under the $SU(6)\times SU(2)$ subgroup
\bea
27 = (6,2) + (\bar{15},1) \\
\bar{27} = (\bar{6},2) + (15,1) \nonumber
\eea

The non-compact generators transforming as $(\bar{6},2)_{nc}$ are  $W_a $ and $\Tilde{W}_a$ ($a=1,2,..,6 $) defined as 
\bea
W_m &=& 12 J^{+}_{m6} \quad , \quad M=1,..,5 \\
W_6 &=& -\sqrt{2} \, \Tilde{N}_{78} = -\frac{1}{\sqrt{2}}\left( \Tilde{M}_{78} + i \Tilde{K}_{78} \right)
\eea
\bea
\Tilde{W}_m &=& \sqrt{2} \Tilde{N}_{m6} = \frac{1}{\sqrt{2}} \left( \Tilde{M}_{m6} + i \Tilde{K}_{m6} \right) \quad . \quad m=1...,5 \\
\Tilde{W}_6 &=&  \frac{1}{2} \left( \Delta + M -  K \right) 
\eea
They satisfy the commutation relations
\bea
[R  , W_a ]=  W_a \quad , \quad [R , \Tilde{W}_a]= \Tilde{W}_a 
\eea
\bea
[J^3 , W_a ] = -\frac{1}{2} W_a \quad , \quad [J^3 , \Tilde{W}_a ] = \frac{1}{2}   \Tilde{W}_a 
\eea
\bea
[J^-,\Tilde{W}_a]= W_a \quad , \quad [J^- , W_a] =0 \quad , \quad [J^+ , \Tilde{W}_a ] = 0  \quad , \quad [J^+ , W_a ] = \Tilde{W}_a 
\eea
\bea
[W_a , W_b ]=0 \quad , \quad [W_a , \Tilde{W}_b ]=0 \quad , \quad [\Tilde{W}_a , \Tilde{W}_b ]=0
\eea
where $a,b,..=1,2,..,6$.

The generators transforming in $(6, 2 )_{nc} $ are given by the Hermitian conjugates of  $W_a $ and $\Tilde{W}_a$  which will be denoted as $W^a = (W_a)^\dagger$ and $\Tilde{W}^a = (\Tilde{W}_a)^\dagger$. They satisfy the commutation relations
 \bea
[R  , W^a ]= - W^a \quad , \quad [R , \Tilde{W}^a]= -\Tilde{W}^a 
\eea
\bea
[J^3 , W^a ] = \frac{1}{2} W^a \quad , \quad [J^3 , \Tilde{W}^a ] = -\frac{1}{2}   \Tilde{W}^a 
\eea
\bea
[J^+,  \Tilde{W}^a]= - W^a \quad , \quad [J^+ , W^a] =0 \quad , \quad [J^- , \Tilde{W}^a ] = 0  \quad , \quad [J^- , W^a ] =- \Tilde{W}^a 
\eea
\bea
[W^a , W^b ]=0 \quad , \quad [W^a , \Tilde{W}^b ]=0 \quad , \quad [\Tilde{W}^a , \Tilde{W}^b ]=0
\eea
\bea
[ W_a , W^b ] =  \left( W^b_a + \delta^b_a ( J^3 -\frac{2}{3} R ) \right) 
\eea
\bea
[ \Tilde{W}_a , \Tilde{W}^b ] =  \left( W^b_a - \delta^b_a ( J^3 +\frac{2}{3} R ) \right) 
\eea
\bea
[ H^-_{mn}, W_p ]=  \sqrt{2}/4 \, \, \epsilon_{mnpqr} \left(M^{qr} - i K^{qr}  \right) = \sqrt{2}/2 \, \, \epsilon_{mnpqr} C^{qr} 
\eea

\bea
[ C^{mn}, \Tilde{W}_p ] = \frac{1}{\sqrt{2}} \delta^m_p J^n_6 -\frac{1}{\sqrt{2}} \delta^n_p J^m_6
\eea
\bea
[ C^{mn},W_a ] = 0
\eea
\bea 
[ C^{mn}, \Tilde{W}_6 ] = \frac{1}{\sqrt{2}} N^{mn} 
\eea
\bea
[J_-^{mn}, W_p ]=\frac{1}{12} ( \delta^m_p  J^n_6 - \delta^n_p J^m_6 )
\eea
\bea
[ H_+^{mn}, W^p ]=  \sqrt{2}/4 \, \, \epsilon^{mnpqr} \left(\Tilde{M}_{qr} + i \Tilde{K}_{qr}  \right) = \sqrt{2}/2 \, \, \epsilon^{mnpqr} \Tilde{C}_{qr} \eea
\bea [H_+^{mn} , W_a]=0 \quad , \quad [H_+^{mn}, W^6 ]= -12 J_-^{mn} 
\eea
\bea
 [ H^-_{mn} , W_6 ] =  12 J^+_{mn} \quad , \quad  [ H^-_{mn} , \Tilde{W}^a ] = 0 \quad , \quad 
  [ H^-_{mn} , \Tilde{W}_p ] = - 6   \epsilon_{mnpqr} J_-^{qr} 
 \eea
 \bea
[ H^-_{mn}, \Tilde{W}_6 ]=  \sqrt{2} \Tilde{C}_{mn}
\eea
\bea [H^{mn}_+ , \Tilde{W}_a ]=0
\eea
\bea
[H_+^{mn}, \Tilde{W}^p ] = 6  \epsilon^{mnpqr} J^+_{qr} \quad , \quad [H_+^{mn}, \Tilde{W}^6 ] = -\sqrt{2} C^{mn} 
\eea
\bea 
[ H^-_{m6} , W_6 ] =0 \quad , \quad [ H^-_{m6} , W_n ] = - 12 J^+_{mn} \quad , m,n=1,,..,5
\eea
\bea
[ H^-_{m6}, \Tilde{W}_6 ]=0 
\eea
where $ m,n,p,..=1,2,..,5 $ and $a,b,..=1,..,6$. 

The unitary lowest weight irreducible representations  of $SO^*(12)$ are uniquely defined by a set of states $|\Omega > $ that transform irreducibly under the maximal compact subgroup $U(6)$  and are annihilated by the lowering operators $H^-_{ab}$ of $SO^*(12)$. Acting with the raising operators $H_+^{ab}$ on the states $|\Omega > $  repeatedly one generates the higher modes of the unitary irrep. 
The lowest weight vector  $ | \psi_{2}^{7/2} \left( x \right);0  \rangle  $ of the minimal unitary representation of $E_{7(-25)}$ is also the lowest weight vector of a unitary irrep of $SO^*(12)$:
\bea
H^-_{ab}  | \psi_{2}^{7/2} \left( x \right);0  \rangle  =0 
\eea 
This state is also a singlet of $SU(2)$ . 
Higher modes of the unitary irrep of $SO^*(12)$  are then generated by the action of raising operators $H_+^{ab}$ 
\bea
 | \psi_{2}^{7/2} \left( x \right);0  \rangle  \quad \bigoplus H_+^{mn} | \psi_{2}^{7/2} \left( x \right);0  \rangle \quad   \bigoplus H_+^{mn} H_+^{pq}  | \psi_{2}^{7/2} \left( x \right);0  \rangle  \bigoplus .....
\eea
To find the decomposition of the minrep of $E_{7(-25)}$  with respect to its $SO^*(12)\times SU(2)$ subgroup we need to identify all the lowest weight irreps of $SO^*(12)\times SU(2)$ in the Hilbert space. These additional lowest weight irreps are generated by the raising operators outside the $SO^*(12)\times SU(2)$ which transform in the $(\bar{6},2)$ representation of $SU(6)\times SU(2)$ namely the generators $W_a $ and $\Tilde{W}_a$ which form a doublet of $SU(2)$. More specifically acting with $W_a$  repeatedly one generates an infinite number of states that are annihilated by the lowering operators $ H^-_{ab} $ of $SO^*(12)$ as well as by $J^-$:
\bea
W_{a_1} W_{a_2} \cdots W_{a_n} | \psi_{2}^{7/2} \left( x \right);0  \rangle 
\eea
They transform irreducibly under the maximal compact subgroup $U(6)$ of $SO^*(12)$ and are   lowest weight irreps of $SU(2)$. They  transform as a symmetric tensor of $SU(6) $ with a definite $R$ charge and eigenvalue of $J_3$ :
\bea
H^-_{ab} W_{a_1} W_{a_2} \cdots W_{a_n} | \psi_{2}^{7/2} \left( x \right);0  \rangle  &=&0 \\
J^- W_{a_1} W_{a_2} \cdots W_{a_n}  | \psi_{2}^{7/2} \left( x \right);0  \rangle  =0 
\eea

\bea 
R \,  W_{a_1} W_{a_2} \cdots W_{a_n} | \psi_{2}^{7/2} \left( x \right);0  \rangle  = (n+6) W_{a_1} W_{a_2} \cdots W_{a_n}  | \psi_{2}^{7/2} \left( x \right);0  \rangle 
\eea
\bea
J^3 W_{a_1} W_{a_2} \cdots W_{a_n} | \psi_{2}^{7/2} \left( x \right);0  \rangle  = -\frac{n}{2} W_{a_1} W_{a_2} \cdots W_{a_n} | \psi_{2}^{7/2} \left( x \right);0  \rangle 
\eea
The fact that $H^-_{ab}$ annihilates the states $W_{a_1} W_{a_2} \cdots W_{a_n} | \psi_{2}^{7/2} \left( x \right);0  \rangle $ follows from the fact that the commutators of $W_a$  and $\tilde{W}_a$ with the lowering operators $H^-_{ab}$ close into compact generators transforming in the $(20,2)$ representation of $SU(6)\times SU(2)$ subgroup of compact $E_6$  that leaves the lowest weight vector  $ | \psi_{2}^{7/2} \left( x \right);0  \rangle $ invariant:
\bea
[ H^-_{ab}, W_c]  | \psi_{2}^{7/2} \left( x \right);0  \rangle  & =&0 \quad , \quad a,b,..=1,..,6 
\eea and satisfy 
\bea
 {[}[ H^-_{ab}, W_c {]},  W_d{]}{]} & =& 0 
\eea
 
 They exhaust the list of all lowest weight irreps of $SO^*(12)\times SU(2)$ inside the minrep of $E_{7(-25)}$. 
Hence we can give the full  decomposition of the minrep of $E_{7(-25)}$ with respect to its subgroup $SO^*(12)\times SU(2)$ readily.
Denoting the Hilbert space of the unitary representations of $SO^*(12)$ whose lowest weight irrep has the $SU(6)$ Dynkin labels $(0,0,0,0,n)$ with $U(1)_R$ charge $(n+6)$ and transforms in the spin $n/2$ representation of $SU(2) $ as $ \mathcal{H}_{n+6} ((0,0,0,0,n);j=\frac{n}{2})$ we have 
\bea
 \mathcal{H}_{minrep}(E_{7(-25)})= \sum_{n=0}^{n=\infty} \bigoplus \mathcal{H}_{n+6}((0,0,0,0,n);j=\frac{n}{2} ) \label{decompE7_SO12}
\eea

As stated earlier the group $SO^*(12)$ is the conformal group of a causal 15 dimensional spacetime coordinatized by the simple Jordan algebra $J_3^{\mathbb{H}}$ of $3\times 3$ Hermitian matrices over the division algebra of quaternions $\mathbb{H}$ . It is a subspace of the exceptional spacetime coordinatized by  $J_3^{\mathbb{O}}$. Its Lorentz group $SU^*(6)$ is a subgroup of the Lorentz group $E_{6(-26)}$ of the exceptional spacetime and we have the decompositions;
\bea
E_{6(-26)} & \supset & SU^*(6)\times SU(2) : \\ \nonumber
78 &=& (35,1) + (1,3) + (20,2) \\ \nonumber
27 &=& (15,1) + (\Tilde{6} ,2) 
\eea
The equation \ref{decompE7_SO12} shows that the minrep of $E_{7(-25)}$ decomposes into the minrep  plus infinitely many different representations  of $SO^*(12)$ that fall into  irreps of an $SU(2)$ subgroup. This $SU(2)$  is the subgroup of the automorphism group $G_2$ of octonions that commutes with the automorphism group  $SU(2)_H$ of its quaternion subalgebra:
\bea
G_2 \supset SU(2)\times SU(2)_H 
\eea

\section{Applications to magical supergravity theories}
$N=2$ Maxwell-Einstein supergravity theories (MESGTs) describe the coupling of vector multiplets to $N=2$ supergravity. In five space-time dimensions $N=2$ MESGTs whose scalar manifolds are symmetric spaces $G/K$ such that $G$ is a symmetry of the Lagrangian are uniquely defined by Euclidean Jordan algebras of degree three.  Their scalar manifolds $\mathcal{M}_5(J)$ are simply the coset spaces
\bea
\mathcal{M}_5(J) =\frac{Lor(J)}{Rot(J)}
\eea
Their scalar manifolds  in four dimensions are
\begin{equation}
 \mathcal{M}_4 = \frac{Conf(J)}{ \widetilde{Lor} (J) \times U(1)}
\end{equation}
where  
$\widetilde{Lor}(J)$ is the compact real form of the Lorentz
group of $J$.  In three dimensions they
lead to scalar manifolds of the form
\begin{equation}
 \mathcal{M}_3 = \frac{QConf(J)}{\widetilde{Conf}(J) \times SU(2)}
\end{equation}
where $QConf(J)$ is the quasiconformal group constructed over the Freudenthal triple system $\mathcal{F}(J)$
of the Jordan algebra $J$ and $\widetilde{Conf}(J)$ is the compact real form of
 $Conf(J)$.
 In Table \ref{scalarmanifolds} we list the scalar manifolds of the magical supergravity theories in $d=3,4,5$ dimensions. 
\begin{table}
\begin{equation} \nonumber
  \begin{array}{|c|c|c|c|} \hline
  ~ &  \mathcal{M}_5 = &  \mathcal{M}_4 = &  \mathcal{M}_3 =  \\
  J & \mathop\mathrm{Lor} \left(J\right)/ \mathop\mathrm{Rot}\left(J\right) &
            \mathop\mathrm{Conf} \left(J\right) / \mathop\mathrm{\widetilde{Lor}}\left(J\right)\times U(1)&
       \mathop\mathrm{QConf} (\mathcal{F}(J))/
                            \widetilde{ \mathop\mathrm{Conf}\left(J\right)} \times \mathrm{SU}(2) \\[7pt]
        \hline
       J_3^\mathbb{R} & \mathrm{SL}(3, \mathbb{R}) / \mathrm{SO}(3) &
                        \mathrm{Sp}(6, \mathbb{R}) / \mathrm{U}(3)  &
            \mathrm{F}_{4(4)} / \mathrm{USp}(6) \times \mathrm{SU}(2)  \\ [7pt]
       J_3^\mathbb{C} & \mathrm{SL}(3, \mathbb{C}) / \mathrm{SU}(3) &
                        \mathrm{SU}(3, 3) / \mathrm{S}\left(\mathrm{U}(3) \times \mathrm{U}(3)\right) &
            \mathrm{E}_{6(2)} / \mathrm{SU}(6) \times \mathrm{SU}(2)   \\ [7pt]
       J_3^\mathbb{H} & \mathrm{SU}^\ast(6) / \mathrm{USp}(6) &
                        \mathrm{SO}^\ast(12) / \mathrm{U}(6) &
            \mathrm{E}_{7(-5)} / \mathrm{SO}(12) \times \mathrm{SU}(2)  \\[7pt]
       J_3^\mathbb{O} & \mathrm{E}_{6(-26)} / \mathrm{F}_4 &
                        \mathrm{E}_{7(-25)} / \mathrm{E}_6 \times \mathrm{U}(1) &
            \mathrm{E}_{8(-24)} / \mathrm{E}_7 \times \mathrm{SU}(2)  \\[7pt]
       \mathbb{R} \oplus \Gamma_n(\mathbb{Q}) &
                      \frac{  \mathrm{SO}(n-1,1) \times \mathrm{SO}(1,1)}{ \mathrm{SO}(n-1)} &
                       \frac{ \mathrm{SO}(n,2) \times \mathrm{SU}(1,1) }{
                             \mathrm{SO}(n) \times \mathrm{SO}(2) \times \mathrm{U}(1)} &
           \frac{ \mathrm{SO}(n+2,4)}{ \mathrm{SO}(n+2) \times \mathrm{SO}(4)} \\ [7pt]
            \hline
  \end{array}
\end{equation}
\caption{\label{scalarmanifolds}
The scalar manifolds $\mathcal{M}_d$ of $N=2$ MESGT's and corresponding 
Jordan algebras $J$ of degree three in $d=3,4,5$ spacetime
dimensions. 
}
\end{table}
The MESGTs defined by simple Jordan algebras $J_3^{\mathbb{A}}$ are called the magical supergravity theories since their symmetry groups in $d=3,4,5$ dimensions are certain noncompact real forms of the groups that appear in the famous Magic Square of Freudental , Rozenfeld and Tits.  The largest magical supergravity theory is defined by the exceptional Jordan algebra $J_3^{\mathbb{O}}$. We shall simply refer to it as the exceptional supergravity. Its global symmetry group in $d=5,4$ and $3$ dimensions are  exceptional group belonging to the E-series , but of different reals forms than maximal supergravity.   The M/superstring theoretic origin of the exceptional supergravity has so far not been established.  For the current status of the searches for an exceptional Calabi-Yau which leads to the exceptional supergravity upon compactification of M/superstring theory we refer to \cite{Gunaydin:2022qvw}. 
\subsection{Exceptional supergravity and composite scenarios }
The exceptional supergravity theory has a common sector with the maximal supergravity, namely the $N=2$ supersymmetric quaternionic magical supergravity theory whose global symmetry group is $SO^*(12)$ in four dimensions. Neither the maximal supergravity nor the exceptional supergravity can accommodate the standard model of particle physics directly since possible gaugings of both theories can not accommodate the standard model gauge group $SU(2)\times U(1)\times SU(3)$. This led the author to suggest that the theory we are looking for might be the one that can accommodate both maximal supergravity and the exceptional supergravity before the so-called first string revolution \cite{Gunaydin:1984be}. 

In the early days of maximal supergravity Murray Gell-Mann argued that to make contact with observation some of the observed particles in Nature may have to be composites of the fields of maximal supergravity.  
First  step in this direction was the proposal by Cremmer and Julia who suggested that  the composite $SU(8)$ local gauge symmetry of maximal supergravity might become  dynamical at the quantum level\cite{Cremmer:1979up}. Authors of \cite{Ellis:1980tf}  extended this proposal and suggested that bound states may form ( bosonic as well as fermionic) that effectively lead to a grand unified theory based on $SU(5)$ which requires the breaking of the family unifying group $SU(8)$. Furthermore the bound states might fall into unitary representations of U-duality group $E_{7(7)}$ which are all infinite dimensional.  Unitary representations of the U-duality groups of extended supergravity theories ($N\geq 4$) using oscillators transforming in the same representation as the vector field strengths and their magnetic duals in four dimensions were given in \cite{Gunaydin:1981dc,Gunaydin:1981yq}. For reviews of the composite scenarios in maximal supergravity  see \cite{Gunaydin:1982gw,Ellis:1983na}.

  Sierra, Townsend and this author  proposed a composite scenario for the exceptional supergravity in which the composite local symmetry $E_6 \times U(1)$ becomes dynamical \cite{Gunaydin:1983rk,Gunaydin:1984be}.  The exceptional group $E_6$ had earlier been proposed as a grand unified gauge group\cite{Gursey:1975ki}. Furthermore in contrast to the U-duality group $E_{7(7)}$ of maximal supergravity the U-duality group $E_{7(-25)}$ of the exceptional supergravity admits unitary lowest weight representations which are intrinsically chiral \cite{Gunaydin:1984be} in the sense that complex representations of the maximal compact subgroup appear in the Hilbert space  without their complex conjugate representations. Hence , in principle, it could accommodate   chiral families of quarks and leptons in a composite scenario of the exceptional supergravity.  
One problem one faces is the fact that unitary irreps of lowest weight type are in general multiplicity free in their K-type decompositions implying that the irrep $27$ of $E_6$ would come with multiplicity one corresponding to one family of quarks and leptons. Our results above show that when $E_6\times U(1)$ is restricted to the $SO(10)\times U(1)\times U(1)$ subgroup the representations of $SO(10)$ appear with infinite multiplicity with  different $U(1)\times U(1)$ charges. In particular $16$ of $SO(10)$ appear with the following charges
\bea
 SO(10)\times U(1)_H \times U(1)_{L_0} : \rightarrow
\sum_{q= 0}^{\infty} (0,0,0,0,1)\times (H=9/2) \times (L_0= q +5/2) \nonumber 
\eea
 If the dynamical group gauge $E_6$ is broken  down to $SO(10)$ subgroup such that masses of composite fermions in different 16 representations get masses proportional to eigenvalues of $L_0$ then one would have an infinite family of fermions of increasing mass. However  the results of \cite{Gunaydin:2007vc} imply that the spectrum will be $N=2$ supersymmetric. Hence one has to break $N=2$ supersymmetry at high energies in order to have truly chiral families of fermions. To make contact with the phenomenology of standard model  one would then have to  find a mechanism to make the fermions beyond three families of quarks and leptons very heavy with $SO(10)$ broken down to the standard model gauge group. 

\subsection{$E_{7(-25)}$ as spectrum generating symmetry group of $5d$ exceptional supergravity }

The orbits of the extremal black holes of  five dimensional $N=2$ MESGTs  defined by Jordan algebras $J$ of degree three that are asymptoticaly flat, static and spherically symmetric were first classified in \cite{Ferrara:1997uz} and studied later in \cite{Ferrara:2006xx,Cerchiai:2010xv}. 
The U-duality group of $5d$ supergravity defined by $J$  is the Lorentz group $Lor(J)$ of $J$ which leaves cubic norm of $J$  invariant.  The entropy of a large  extremal black hole in these  theories is determined  by the norm form of $Q\in J$ representing the charges of the black hole and hence are invariant under $Lor(J)$.  However, one finds that the small extremal black holes have a larger symmetry, since their entropy vanish. This larger group is the semidirect product group $\left( Lor(J) \times SO(1,1)\right)\rtimes S_J $, where $S_J$ is the Abelian  group of special conformal transformations of $J$. However under the full conformal group $Conf(J)$ the lightlike elements of $J$ do not remain lightlike in general.  
 Conformal group $Conf{J}$ changes the norm  $Q\in J$  and hence the corresponding entropy of extremal  black hole
whose charges are given by  $Q$.
Hence  conformal groups $Conf(J)$ of Jordan algebras that define the $5d$ MESGTs
 were proposed as spectrum generating
symmetry groups of these theories~\cite{Ferrara:1997uz,Gunaydin:2000xr,Gunaydin:2004ku,Gunaydin:2005gd,Gunaydin:2009pk}.
 U-duality group of $4d$ MESGT defined by $J$ is isomorphic to the conformal group $Conf(J)$ and acts linearly on the electric and magnetic charges of an extremal black hole represented by the elements of a Freudenthal triple system defined $\mathcal{F}(J)$ over $J$\cite{Gunaydin:1983bi,Gunaydin:2000xr,Ferrara:1997uz,Gunaydin:2005gd,Gunaydin:2005zz,Gunaydin:2009pk}.

Even though we do not yet understand fully the M/superstring theoretic origin of the exceptional supergravity we expect its quantum completion to break its U-duality group down to its arithmetic subgroup as well as its spectrum generating extension. For the $5d$ exceptional supergravity they are $E_{6(-26)}(\mathbb{Z}) $ and $E_{7(-25)}(\mathbb{Z})$, respectively. 
Extremal black hole charges are then represented by the elements of exceptional Jordan algebra taken over the integral octonions $\mathcal{R}$ of Coxeter which we denote as $J_3^{\mathbb{O}}(\mathcal{R})$ \cite{Gunaydin:2022qvw}. Mathematicians have studied the action of arithmetic  $E_{7(-25)}(\mathbb{Z})$ on $J_3^{\mathbb{O}}(\mathcal{R})$  and defined some important   holomorphic modular forms, in particular the singular modular form $E_4(Z)$  \cite{Kim}. Furthermore in a remarkable paper Elkies and Gross calculated the number  of linearly independent charge configurations of rank one elements of  $J_3^{\mathbb{O}}(\mathcal{R})$ with a given trace\cite{ElkiesGross}.  This result was extended to rank two elements with a given trace and a spur form in \cite{Krieg,MR1195510}. Rank one and rank two elements of $J_3^{\mathbb{O}}$ describe small extremal black holes in $5d$ exceptional supergravity and their orbits under the continuous U-duality group $E_{6(-26)}$ were calculated in \cite{Ferrara:1997uz,Ferrara:2006xx}.  Rank one elements of $J_3^{\mathbb{O}} $ correspond to points in Moufang plane and to  pure states in octonionic quantum mechanics parametrized by the coset space $F_4/SO(9)$\cite{Gunaydin:1978jq}. While working over the integral octonions of Coxeter one has to count the degeneracy of a rank one element with a given trace as was done by Elkies and Gross. Since the restriction to discrete arithmetic subgroup arises at the quantum level Kidambi and this author (GK) defined the quantum degeneracy of a rank one black hole to be the degeneracy calculated by Elkies and Gross\cite{Gunaydin:2022qvw}.\footnote{Note that this definition of quantum degeneracy is purely number theoretic and  is not to be confused with the degeneracy  of microstates of stringy black holes!}  Using the results of Elkies and Gross,   GK showed that the quantum degeneracies of charge configurations of rank one BPS black holes in the exceptional supergravity are given by the Fourier coefficients of the unique singular modular form $E_4(Z)$ of weight 4 of $E_{7(-25)}(\mathbb{Z})$ . Similarly the quantum degeneracies of rank two BPS black holes are given the Fourier coefficients of the unique singular modular form $E_8 (Z) = (E_4(Z))^2 $ of weight 8. 

The minimal unitary representation of $E_{7(-25)}$ studied above has Gelfand-Kirillov dimension 17 corresponding to 16 variables in the 16 of $SO(10)$ and a singlet. These variables when represented as elements of the exceptional spacetime coordinatized by elements of  $J_3^{\mathbb{O}}$ 
have the form
\[ \begin{pmatrix} 0&0& o_1 \\
0&0& o_2 \\
\bar{o}_1 & \bar{o}_2 & \rho 
\end{pmatrix} 
\]
where $o_1$ and $o_2$ are octonions and $\rho$ is the singlet coordinate. 
Hence they correspond to rank two elements in general which can be restricted to rank one element by setting the $\rho$ equal  to zero. This suggests  that the singular modular form $E_4(Z)$ as well as $E_8(Z)$ must admit uplifts  to automorphic forms associated with the minrep of $E_{7(-25)}$ which we hope to return to in a separate study.\footnote{ For a clear exposition of the extensions of modular forms to automorphic forms associated with unitary lowest weight representations we refer to the book \cite{Fleig:2015vky}. }

{\bf Acknowledgements} : I would like to thank Sudarshan Fernando, Karan Govil, Axel Kleinschmidt, Kilian Koepsell , Hermann Nicolai and Oleksandr Pavlyk for enjoyable collaborations and many stimulating discussions on minimal unitary representations over the years.

\section{Appendix A} 
\label{appendixA}

{\bf  A review of 
the Minimal Unitary Realization of $E_{8(-24)}$ } 

The minimal unitary realization of $ \mathfrak{e}_{8(-24)}$ as a quasiconformal Lie algebra  was given in \cite{Gunaydin:2004md} in  $SU^\ast(8)$ as well as $SU(6,2)$  bases  in terms of 28 bosonic annihilation and creation operators transforming covariantly under the $SU^*(8)$ and $SU(6,2)$ subgroups of $E_{7(-25)}$ , respectively. We shall briefly review the construction in $SU(6,2)$ basis in this section. 
The oscillators satisfy the commutation relations
\begin{equation}
  \left[ \Tilde{Z}^{ab} \,, Z^{cd} \right] = \frac{1}{2} \left( \eta^{ca}\eta^{db} - \eta^{cb}\eta^{da} \right) \,.
\end{equation}
where $Z^{ab}=-Z^{ba} $ and $\Tilde{Z}^{ab} =- \Tilde{Z}^{ba} $ and $a,b,..=1,2,..,8$ and  the  Hermiticity conditions
\begin{equation}
    \left(Z^{ab}\right)^\dagger = \Tilde{Z}^{ab} = \eta^{ac}\eta^{bd} \Tilde{Z}_{cd}
\end{equation}
$\eta$ is the $SU(6,2)$  invariant metric
  $\eta = \mathop\mathrm{Diag} \left( 1,1,1,1,1,-1,1,-1\right)$ with which the indices are raised and lowered\footnote{Note that our labelling of indices correspond to interchanging the indices 6 and 7 in \cite{Gunaydin:2004md}.} . The   28  creation operators in this basis are $Z^{mn} $  , $Z^{68}$ and $ \Tilde{Z}_{mi}$ with $m,n=1,2,3,4,5,7$ and $j=6,8$.
Generators of $\mathfrak{e}_{7(-25)}$ subalgebra of $ \mathfrak{e}_{8(-24)}$ in $SU(6,2)$ covariant basis take the form
\begin{equation}
\begin{split}
   {J^{a}}_b &= 2 {Z}^{ac} \Tilde{Z}_{bc} - \frac{1}{4} {\delta^a}_b {Z}^{cd} \Tilde{Z}_{cd} \\
   J^{abcd} &= \frac{1}{2} {Z}^{[ab} {Z}^{cd]} -
                  \frac{1}{48} \epsilon^{abcdefgh} \Tilde{Z}_{ef} \Tilde{Z}_{gh}
\end{split}
\end{equation}
with Hermiticity conditions
\begin{equation}
   \left( {J^{a}}_b \right)^\dagger = \eta^{ad} \eta_{bc} {J^c}_d \qquad
   \left( J^{abcd} \right)^\dagger = - \frac{1}{24} {\epsilon^{abcd}}_{efgh} J^{efgh}
\end{equation}
They satisfy the commutation relations
\begin{equation}
\begin{split}
  \left[ {J^a}_b \,, {J^c}_d \right] &= {\delta^c}_b {J^a}_d - {\delta^a}_d {J^c}_b \\
  \left[ {J^a}_b \,, J^{cdef} \right] &= -4 {\delta^{[c}}_b J^{def]a} - \frac{1}{2} {\delta^a}_b J^{cdef} \\
  \left[ J^{abcd} \,, J^{efgh} \right] &=  \frac{1}{36} \epsilon^{abcdp[efg} {J^{h]}}_p
\end{split}
\end{equation}
and the  quadratic Casimir of $\mathfrak{e}_{7(-25)}$ takes the form
\begin{equation}
\begin{split}
   \mathcal{C}_2 \left( \mathfrak{e}_7 \right) &=  \frac{1}{6} {J^a}_b {J^b}_a + J^{abcd} \left( \epsilon J \right)_{abcd}
        \\
     &  = \frac{1}{2} \left( \Tilde{Z}_{ab} Z^{bc} \Tilde{Z}_{cd} Z^{da} +
                     Z^{ab} \Tilde{Z}_{bc} Z^{cd} \Tilde{Z}_{da} \right)\\
       &  - \frac{1}{8} \left( \Tilde{Z}_{ab} Z^{ab} \Tilde{Z}_{cd} Z^{cd} +
                  Z^{ab} \Tilde{Z}_{ab}  Z^{cd} \Tilde{Z}_{cd}  \right) + 14 \\
      &  \mspace{10mu} + \frac{1}{96} \epsilon_{abcdefgh} Z^{ab} Z^{cd} Z^{ef} Z^{gh}
        + \frac{1}{96} \epsilon^{abcdefgh} \Tilde{Z}_{ab}
                          \Tilde{Z}_{cd} \Tilde{Z}_{ef} \Tilde{Z}_{gh}
\end{split}
\end{equation}
The quartic invariant operator $I_4$ of  $\mathfrak{e}_{7(-25)}$ is related to the Casimir operator as \cite{Gunaydin:2004md}
\eq \label{quartic-e7}
  I_4 \left( Z \,, \Tilde{Z} \right)= \mathcal{C}_2 \left( \mathfrak{e}_{7(-25)}\right)+ \frac{323}{16}
  \en
The  grade $-2$ and $-1$ generators of $\mathfrak{e}_{8(-24)}$ are given by
\begin{equation}
  E = \frac{1}{2} x^2  \qquad E^{ab} = x Z^{ab} \qquad \qquad \Tilde{E}_{ab} = x \Tilde{Z}_{ab}
  \label{eq:su62negativegrade}
\end{equation}
Generators in grade $+1$ subspace $\mathfrak{g}^{+1}$ are obtained by commutation of  grade -1 generators $E^{ab}$ and $\Tilde{E}_{ab}$   with grade +2 generator $F$ given by
\begin{equation}
   F = \frac{1}{2} p^2 + 2 x^{-2} I_4
\end{equation}

  We have
\begin{equation}
\begin{split}
   F^{ab} & = i \left[ E^{ab}\,, F \right] =  - p Z^{ab} + 2 i x^{-1} \left[ Z^{ab} \,, I_4 \right] \\
   \Tilde{F}_{ab} & = i \left[ \Tilde{E}_{ab}\,, F \right] =
          - p \Tilde{Z}_{ab} + 2 i x^{-1} \left[\Tilde{Z}_{ab} \,, I_4 \right]
\end{split}
\label{eq:su62grade1}
\end{equation}
Explicitly we have
\begin{equation}
\begin{split}
   F^{ab} = &- p Z^{ab} - \frac{i}{12} x^{-1} \epsilon^{abcdefgh} \Tilde{Z}_{cd} \Tilde{Z}_{ef} \Tilde{Z}_{gh} \\
       & + 4 i x^{-1} Z^{c[a}\Tilde{Z}_{cd} Z^{b]d} + \frac{i}{2} \, x^{-1} \left( Z^{ab} \Tilde{Z}_{cd} Z^{cd} +
       Z^{cd}  \Tilde{Z}_{cd} Z^{ab}\right) \\
  F_{ab} = & -p \Tilde{Z}_{ab} + \frac{i}{12} x^{-1} \epsilon_{abcdefgh} Z^{cd} Z^{ef} Z^{gh} \\
        & - 4 i x^{-1} \Tilde{Z}_{c[a} Z^{cd} Z_{b]d} - \frac{i}{2}\, x^{-1} \left( \Tilde{Z}_{ab} Z^{cd} \Tilde{Z}_{cd}
        + \Tilde{Z}_{cd} Z^{cd} \Tilde{Z}_{ab}  \right)
\end{split}
\end{equation}

They satisfy the commutation relations
\begin{equation}
\begin{array}{c}
   \left[ E, F \right] = - \Delta \\[6pt]
\begin{aligned}
  \left[ \Delta,\, F \right] &= 2 F \cr
  \left[ \Delta,\, F^{ab} \right] &= F^{ab} \cr
  \left[ \Delta,\, \Tilde{F}_{ab} \right] &= \Tilde{F}_{ab}
\end{aligned}
 \quad
\begin{aligned}
  \left[ \Delta,\, E \right] &= - 2 E \cr
  \left[ \Delta,\, E^{ab} \right] &= - E^{ab} \cr
  \left[ \Delta,\, \Tilde{E}_{ab} \right] &= - \Tilde{E}_{ab}
\end{aligned}
\end{array}
 \quad
\begin{aligned}
   \left[ E \,, F^{ab} \right] &= - i E^{ab} \cr
   \left[ E \,, \Tilde{F}_{ab} \right] &= - i \Tilde{E}_{ab} \cr
   \left[ F \,, E^{ab} \right] &=  i F^{ab} \cr
   \left[ F \,, \Tilde{E}_{ab} \right] &=  i \Tilde{F}_{ab}
\end{aligned}
\end{equation}
\begin{equation}
\begin{aligned}
  \left[ E,\, E^{ab} \right] &= 0 \cr
  \left[ F,\, F^{ab} \right] &= 0 \cr
\end{aligned}
\quad
\begin{aligned}
  \left[ E,\, \Tilde{E}_{ab} \right] &= 0 \cr
  \left[ F,\, \Tilde{F}_{ab} \right] &= 0
\end{aligned}
\quad
\begin{aligned}
  \left[ \Tilde{E}_{ab} \,, {E}^{cd} \right] &= 2 \, \delta^{cd}_{ab} \, E \cr
   \left[ \Tilde{F}_{ab} \,, {F}^{cd} \right] &= 2 \, \delta^{cd}_{ab} \, F
\end{aligned}
\end{equation}
\begin{equation}
\begin{aligned}
   \left[ E^{ab} \,, F^{cd} \right] &= - 12 i J^{abcd} \cr
   \left[ \Tilde{E}_{ab} ,\, \Tilde{F}_{cd} \right] &=  12 i \left( \epsilon J\right)_{abcd}
\end{aligned}
\quad
\begin{aligned}
   \left[ \Tilde{E}_{ab} ,\, F^{cd} \right] &= - 4 i {\delta^{[c}}_{[a} {J^{d]}}_{b]}
          - i \delta^{cd}_{ab} \Delta  \cr
   \left[ E^{ab} ,\, \Tilde{F}_{cd} \right] &= - 4 i {\delta^{[a}}_{[c} {J^{b]}}_{d]}
          + i \delta^{ab}_{cd} \Delta  \cr
\end{aligned}
\end{equation}
\begin{equation*}
  \begin{aligned}
  \left[ {J^a}_b \,, E^{cd} \right] & = {\delta^c}_b E^{ad} + {\delta^d}_b E^{ca} - \frac{1}{4} {\delta^a}_b E^{cd}  \cr
  \left[  \Tilde{E}_{cd}  \,, {J^a}_b \right] & = {\delta^a}_c \Tilde{E}_{bd} + {\delta^a}_d \Tilde{E}_{cb} -
                                                \frac{1}{4} {\delta^a}_b \Tilde{E}_{cd}
  \end{aligned}
\end{equation*}
\begin{equation*}
\begin{aligned}
   \left[ J^{abcd} \,, \Tilde{E}_{ef} \right] &=  \delta^{[ab}_{ef} E^{cd]} \cr
   \left[ J^{abcd} \,, E^{ef} \right] &=  \frac{-1}{24} \epsilon^{abcdefgh} \Tilde{E}_{gh}
\end{aligned}
\end{equation*}

The quadratic Casimir of $\mathfrak{e}_{8(-24)}$ in the above basis takes the form
\begin{equation}
\begin{split}
  \mathcal{C}_2 &= \frac{1}{6} \,{J^a}_b {J^b}_a + J^{abcd} \left( \epsilon J\right)_{abcd}
    + \frac{1}{6} \left( E F + F E +\frac{1}{2} \,\Delta^2 \right) \\
   & -
   \frac{i}{12} \left(  \Tilde{F}_{ab} E^{ab} +  E^{ab} \Tilde{F}_{ab}- \Tilde{E}_{ab} F^{ab} - F^{ab} \Tilde{E}_{ab}
   \right)
\end{split}
\end{equation}
and reduces to a c-number 
\begin{equation} 
   \mathcal{C}_2 \left( \mathfrak{e}_{8(-24)} \right) = - 40 \,.
\end{equation}

\section{Appendix B}
\label{appendixB}
{\bf  $U(5)$ covariant formulation of  the representations of $SO(10)$ subgroup of $SO(10,2) $ }

The $U(5)$ generators are given by
\bea
J^m_n &= &2 Z^{ma} \Tilde{Z}_{na} -1/3 \delta^m_n ( Z^{ab} \Tilde{Z}_{ab} ) \\    
&= &2 Z^{mp} \Tilde{Z}_{np} + 2 Z^{m6} \Tilde{Z}_{n6} -1/3 \delta^m_n ( Z^{pq}\Tilde{Z}_{pq} + 2 Z^{p6} \Tilde{Z}_{p6} ) \\
&=&\hat{J}^m_n - \frac{1}{5} \delta^m_n ( J^6_6) 
\eea
where $a,b,..=1,..,6$ and $m,n,..=1,..,5$ and $\hat{J}^m_n $ denotes the generators of $SU(5)$
\bea
\hat{J}^m_n &= &2 Z^{mp} \Tilde{Z}_{np} + 2 Z^{m6} \Tilde{Z}_{n6} -2/5 \delta^m_n ( Z^{pq} \Tilde{Z}_{pq} + 2 Z^{p6} \Tilde{Z}_{p6}) \\
J^6_6 &=& -1/3 Z^{mn} \Tilde{Z}_{mn} +4/3 Z^{m6} \Tilde{Z}_{m6}
\eea
The other generators  of the maximal compact subgroup $SO(10)$ of $SO(10,2)$ are given by
\bea
J_-^{mn} = \frac{1}{6} Z^{mn} Z^{78} -\frac{1}{12} \epsilon^{mnpqr} \Tilde{Z}_{pq} \Tilde{Z}_{r6}  \\
J^+_{mn} = -\frac{1}{6} \Tilde{Z}_{mn} \Tilde{Z}_{78} + \frac{1}{12} \epsilon_{mnpqr} Z^{pq} Z^{r6} 
\eea
They satisfy the commutation relations
\bea
[J^+_{mn},J_-^{pq}]= \frac{1}{144} \left( \delta_n^p J^q_m -\delta^p_m J^q_n - \delta^q_n J^p_m + \delta^q_m J^p_n  \right) - \frac{1}{216} \left( \delta^p_m \delta^q_n - \delta^p_n \delta^q_m \right) U 
\eea
where
\[ U= -\frac{1}{4} \sum_{a,b=1}^6 \left(Z^{ab} \Tilde{Z}_{ab} + \Tilde{Z}_{ab} Z^{ab} \right) +\frac{3}{2} \left(Z^{78} \Tilde{Z}_{78} +\Tilde{Z}_{78}  Z^{78} \right) \]
We should note that the Hermiticity conditions we impose imply that $Z^{mn}, \Tilde{Z}_{m6}$ and $\Tilde{Z}_{78}$ are creation operators and $\Tilde{Z}_{mn} , Z^{m6}$ and $Z^{78}$ are annihilation operators. Hence the compact $SO(10)$ generators above involve one annihilation and one creation operator as expected. The Fock space is spanned by states created by the action of the creation operators which form representations of $U(5)$. The Lie algebra of $SO(10)$ admits  a  three-grading with respect to its $U(5)$ subalgebra determined  by $Q$ 
\bea
\mathfrak{so}(10) &=J_-^{mn}  \oplus ( \hat{J}^m_n +Q ) \oplus J^+_{mn} \\
\left[Q , J^+_{mn} \right]&= 4 J^+_{mn}  \qquad , 
\left[ Q , J_-^{mn} \right] =-4 J_-^{mn} 
\eea
where $Q= 36 [J^+_{mn}, J_-^{mn}] = Z^{mn}\Tilde{Z}_{mn} + 6 \Tilde{Z}_{m6} Z^{m6} -10 \Tilde{Z}_{78} Z^{78} $.  

Let  $|\Omega_+ \rangle$ denote a set of states that  transform irreducibly under $U(5)$ and are annihilated by the raising operators $J^+_{mn}$. Then by acting  with the lowering operators $J_-^{mn}$  repeatedly on  $|\Omega_+ \rangle$  one generates a set of states that form an irreducible representation of $SO(10)$\cite{Gunaydin:1988kz}. :
\bea 
J^+_{mn} |\Omega_+ \rangle =0 \Longrightarrow \{ |\Omega_+ \rangle \oplus J_-^{mn}|\Omega_+ \rangle \oplus  J_-^{pq} \,  J_-^{mn}|\Omega_+ \oplus .... \} \Longleftrightarrow  \text{ irrep} \,\,\, \mathfrak{so}(10)
\eea
Every unitary irrep of a compact group  has an highest weight as well as a lowest weight vector. Therefore for every set of states $|\Omega_+ \rangle$ there exist another set of states $|\Omega_- \rangle$ annihilated by the lowering  operators and transforming irreducibly under $U(5)$ subalgebra such that by acting on $|\Omega_- \rangle$ with raising operators one can  generate the same irrep of $SO(10)$\cite{Gunaydin:1988kz}:
\bea 
J^{mn}_- |\Omega_- \rangle =0 \Longrightarrow \{ |\Omega_- \rangle \oplus J^+_{mn}|\Omega_- \rangle \oplus  J^+_{pq} \,  J^+_{mn}|\Omega_- \oplus .... \} \Longleftrightarrow \text{ irrep} \,\,\, \mathfrak{so}(10)
\eea

\textit{ By abuse of standard terminology we shall refer to the states $|\Omega_+ \rangle $ and $ |\Omega_- \rangle $ annihilated by the raising and lowering operators  $J^+_{mn}$ and $ J_-^{mn}$ , respectively, as highest and lowest weight irreps. If they are singlets we shall simply call them highest or lowest weight vectors. } The highest weight irreps $|\Omega_+ \rangle $  are of the form 
\bea 
|\Omega_+ \rangle  \Longrightarrow   \Tilde{Z}_{m6} \Tilde{Z}_{n6} \cdots \Tilde{Z}_{p6} |0\rangle \label{u5hwv}
\eea
The corresponding lowest weight vectors are of the form
\bea 
|\Omega_- \rangle  \Longrightarrow   \Tilde{Z}_{78} \Tilde{Z}_{78} \cdots \Tilde{Z}_{78} |0\rangle \label{u5lwv}
\eea
The lowest weight state  $ \Tilde{Z}_{78} |0\rangle $ leads to the spinor representation $16$ of $SO(10)$ and its  highest weight irrep is  $\Tilde{Z}_{p6} |0\rangle$. The spinor irrep $16$ of  $SO(10)$ decomposes as follows with respect to its $SU(5)\times U(1)_Q$ subgroup
\bea
16 = 1_{-5} + \bar{5}_{3} + 10_{-1}  
\eea
where the subscript refers to the $U(1)_Q$ charge $Q$. The $SU(5) \times U(1)_Q$  Dynkin labels of the highest weight irreps of the form \ref{u5hwv}  are $(n,0,0,0)_{3n}$ and the Dynkin labels of the corresponding lowest weight states \ref{u5lwv} are  $(0,0,0,0)_{-5n}$ where the subscript denotes the $U(1)_Q$ charges. They lead to the  irreps of $SO(10)$ with the Dynkin labels $(0,0,0,0,n)$:
\bea \label{so10irreps}
|\Omega_+ \rangle & = & \Tilde{Z}_{m6} \Tilde{Z}_{n6} \cdots \Tilde{Z}_{p6} |0\rangle  \Longleftrightarrow |(n,0,0,0)_{5n} \rangle \,\, \textrm{of \, U(5) }\Longleftrightarrow |(0,0,0,0,n)\rangle  \textrm{ of SO(10)} \nonumber \\
|\Omega_- \rangle &=&  \left(\Tilde{Z}_{78}\right)^n |0\rangle  \Longleftrightarrow  |(0,0,0,0)_{-5n}\rangle  \,\, \textrm{of \, U(5) }\Longleftrightarrow |(0,0,0,0,n)\rangle  \textrm{ of SO(10)} 
\eea

The decomposition of the next irrep $(126)= (0,0,0,0,2)$  of $SO(10)$ with respect to $SU(5)\times U(1)_Q$ subgroup is 
\bea 
(0,0,0,0,2) = 126 = 1_{(-10)}+\bar{5}_{(-2)}+ 10_{(-6)}+\bar{15}_{(6)}+45_{(2)}+\bar{50}_{(-2)} 
\eea
and that of irrep $\overline{672}=(0,0,0,0,3)$ is
\bea
(0,0,0,0,3)&=&  1_{(-15)} + \bar{5}_{(-7)}+10_{(-11)} + \bar{15}_{(1)} + 35_{(9)} +45_{(-3)} +\bar{50}_{(-7)}\\ \nonumber && + \overline{126}_{(5)}+ 175''_{(-3)} +\overline{210}_{(1)} 
\eea  
The noncompact generators of $SO(10,2) $ are $J^m_6 \, , J^6_m \, , J^{m6}_- $ and $J^+_{m6}$. Of these the generators $J^6_n$ and $J_-^{m6}$ involve di-annihilation operators and $ J^m_6$ and $J^+_{m6}$ involve di-creation operators. They satisfy

\bea [Q, J^m_6] =-2 J^m_6 \qquad , \,  [Q, J^{m6}_-]= - 2 J^{m6}_-  \eea
\bea
[Q, J^6_m] &= 2 J^6_m \qquad , \,  [Q,  J_{m6}^+] = 2 J_{m6}^+ 
\eea

\section{Appendix C}
\label{appendixC}
{\bf Compact subalgebra $\mathfrak{e}_6 $ of $\mathfrak{e}_{7(-25)}$ } \\

The grade zero compact subalgebra $\mathfrak{e_{6}} $ of the Lie algebra $\mathfrak{e_{7(-25)}} $ admits three grading with respect to its subalgebra $\mathfrak{so}(10) + \mathfrak{u}(1)_F $ generated by $S_{ij} (i,j =1,\cdots ,10)$ and the $U_F$ generator $F= 2/3 (S_{(11,12)}-2 L_0)$:
\bea
\mathfrak{e_{6} } &=&  \mathfrak{k^{-1}} \oplus  \mathfrak{so}(10) \oplus \mathfrak{u}(1)_F \oplus 
\mathfrak{k^{+1}} \\
78 &=& \bar{16} \oplus ( 1+ 45) \oplus 16  
\eea 

\bea
[ F, \mathfrak{k}^{\pm} ] =\pm \mathfrak{k}^{\pm} 
\eea
The generators in the grade $\pm 1$ subspaces are  as follows
\bea
\mathfrak{k^{-1}} =(\Tilde{C}_{mn} \oplus C^{m6}  \oplus C^{78})  \\
 \mathfrak{k^{+1}} =(C^{mn} \oplus \Tilde{C}_{m6} \oplus \Tilde{C}_{78})  
 \eea
 where
 \[\Tilde{C}_{mn}=1/2 (\Tilde{M}_{mn} + i \Tilde{K}_{mn} )\quad , \quad C^{m6}=1/2 (M^{m6} + i K^{m6} )\quad ,\quad
 C^{78}=1/2(M^{78} + i K^{78}) \] \[ \quad C^{mn}=1/2 (M^{mn} - i K^{mn} ) \quad , \quad 
\Tilde{C}_{m6}=1/2((\Tilde{M}_{m6} - i \Tilde{K}_{m6} ) \quad , \quad  \Tilde{C}_{78}=1/2(\Tilde{M}_{78} - i \Tilde{K}_{78})
 \]

 They satisfy the commutation relations
 \bea
 [C^{mn} ,\Tilde{C}_{pq} ]& = & 1/2 ( \delta^m_p J^n_q +  \delta^n_q J^m_p -  \delta^m_q J^n_p -  \delta^n_p J^m_q ) \\ \nonumber
  && +1/4 (\delta^m_q \delta^n_p - \delta^m_p \delta^n_q) ( M + K +\frac{2}{3} U ) 
 \eea
 \bea
 [C^{m6} ,\Tilde{C}_{p6}  ] = - 1/2 \delta^m_p J^6_6  - 1/2  J^m_p  -1/4 \delta^m_p ( M + K -\frac{2}{3} U ) 
 \eea
 \bea
 [C^{78} ,\Tilde{C}_{78} ]=-1/4 (M+K+2 U)
 \eea
 \bea
 [ C^{m6} , \Tilde{C}_{78} ] ]=0  \quad , \quad   [ C^{m6} , C^{78}]=0   
 \eea
 \bea
 [ C^{mn} , C^{p6} ] = 3 \epsilon^{mnpqr} J^+_{qr} 
 \eea 
 \bea
 [\Tilde{C}_{mn} , \Tilde{C}_{p6} ] = 3 \epsilon_{mnpqr} J_-^{qr} 
 \eea 
  \bea
 [ C^{mn} , \Tilde{C}_{p6} ] = 0 
 \eea 
  \bea
 [ C^{mn} , \Tilde{C}_{78} ] = 0 
 \eea 
 \bea
 [ C^{mn} , C^{78} ] = 6 J_{-}^{mn}
 \eea
 \bea
 [ \Tilde{C}_{mn} , \Tilde{C}_{78} ] = 6 J^{+}_{mn}
 \eea

 The non-compact generators in grade $\pm 1$ subspaces of $\mathfrak{e_{7(-25)}} $ given in \ref{comp3grade} carry the following $U(1)_F$ charges
 \bea
 [ F, S_i^- ] &=&-2/3 S_i^- \, , \quad [F, \Tilde{N}_{mn} ] = 1/3 \Tilde{N}_{mn} \, , \quad [F, N^{m6} ] = 1/3 N^{m6}  
 \eea
 \bea
  [F, N^{78} ] = 1/3 N^{78}  \, , \quad [F, L^-] =4/3 L^- \nonumber 
 \eea
 \bea
 [ F, S_i^+ ] &=& 2/3 S_i^+ \, , \quad [F, N^{mn} ] = -1/3 N^{mn} \, , \quad [F, \Tilde{N}_{m6} ] = - 1/3\Tilde{N}_{m6}  
 \eea
 \bea
  [F, \Tilde{N}_{78} ] = - 1/3 \Tilde{N}_{78}  \, , \quad [F, L^+] =- 4/3 L^+ \nonumber 
 \eea
 The sum over the $U(1)_F$ charges of grade $+1$ or grade $-1$ generators of $\mathfrak{e_{7(-25)}} $ add up to zero as they must. 
 The Casimir operator of the compact $E_6$ is given by
 \bea
 C_2(E_6) &=& \frac{1}{12}  S_{ij}S_{ij} +\frac{1}{8} F^2 + \frac{1}{6} ( \Tilde{C}_{mn} C^{mn} + C^{mn} \Tilde{C}_{mn} ) \nonumber \\
 && -\frac{1}{3} ( \Tilde{C}_{m6} C^{m6} + C^{m6} \Tilde{C}_{m6} ) -\frac{1}{3}(\Tilde{C}_{78} C^{78} + C^{78} \Tilde{C}_{78} )\nonumber \\
 && =  \frac{1}{12}  S_{ij}S_{ij}  + \frac{1}{6} ( \Tilde{C}_{mn} C^{mn} + C^{mn} \Tilde{C}_{mn} )+\frac{1}{8} F^2 \nonumber \\
 && -\frac{1}{3} ( \Tilde{C}_{m6} C^{m6} + C^{m6} \Tilde{C}_{m6} ) -\frac{1}{3}(\Tilde{C}_{78} C^{78} + C^{78} \Tilde{C}_{78} )
 \eea 
 where $i,j,..=1,..,10$ and $m,n,..=1,..,5$. We note that  
 \[  \frac{1}{6} (S_{11,12})^2+\frac{1}{3} (L_0)^2 = \frac{1}{18}(S_{11,12}- 2L_0)^2 +\frac{1}{9} ( S_{11,12} +L_0)^2 =\frac{1}{8} F^2 + \frac{1}{9} R^2\]
 where  $ \frac{1}{9} R^2 $ is the quadratic Casimir of $U(1)_R$.
 The commutation relations between the compact generators  in grade $+1$ subspace $\mathfrak{k}^{+1}$ of $\mathfrak{e}_6$ and the noncompact generators in grade $+1$ subspace $\mathfrak{g}^{+1}$ of $\mathfrak{e}_{7(-25)}$, both in $16$ of $SO(10)$, are
 \bea
 [ C^{mn}, N^{pq} ]&=& 6 \epsilon^{mnpqr} J^+_{r6}  \quad , \quad [C^{mn}, \Tilde{N}_{p6} ]= \frac{1}{2} ( \delta^m_p J^n_6 - \delta^n_p J^m_6 ) \\
 {[} C^{mn} , \Tilde{N}_{78} {]}&=&0 \quad , \quad {[} \Tilde{C}_{m6} , N^{pq} {]} = 0 \quad ,\quad  {[} \Tilde{C}_{m6}, \Tilde{N}_{78} {]} = - 6 J^+_{m6} \quad , 
 {[} \Tilde{C}_{78}, \Tilde{N}_{78} {]} =0 \nonumber
 \eea
 On the other hand the commutators of compact generators in $16$ and $\bar{16}$ of $SO(10)$ and noncompact generators in $\bar{16}$ and $16$ of $SO(10)$ close into $SU(1,1) $ generators $L^{-}$ and $L^+$, respectively :
 \bea
 [ C^{mn} , \Tilde{N}_{pq} ] = -\frac{1}{2} \delta^{mn}_{pq} L^- \quad , \quad [ \Tilde{C}_{mn} ,N^{pq} ] = \frac{1}{2} \delta^{mn}_{pq} L^+ \\
 {[} \Tilde{C}_{m6} , N^{n6} {]} = \frac{1}{2} \delta_m^n L^- \quad , \quad 
 {[} C^{m6}, \Tilde{N}_{n6} {]} = - \frac{1}{2} \delta^m_n L^+  \\
  {[} \Tilde{C}_{78} , N^{78} {]} = \frac{1}{2} L^- \quad , \quad 
   {[} C^{78}, \Tilde{N}_{78} {]} = - \frac{1}{2}  L^+
 \eea
 \section{ Appendix D}
 \label{appendixD}
 {\bf Truncations of the minrep for $E_{7(-25)}$  to subgroups related to lower dimensional critical spacetimes}

  The results for the minrep of $E_{7(-25)}$can be truncated to minimal unitary representations of the conformal groups of Jordan algebras $J_3^{\mathbb{A}}$ for $\mathbb{A}= \mathbb{H}, \mathbb{C}$ or $\mathbb{R}$ which have the Minkowskian conformal  groups $SO(d,2)$ in $d=6,4$ and 3 dimensions as subgroups. These groups are $SO^*(12), SU(3,3)$ and $Sp(6,\mathbb{R})$ , respectively. They correspond to the decompositions
\bea
   E_{7(-25)} \supset SO^*(12) \times SU(2) \\
   SO^*(12) \supset SU(3,3) \times U(1) \\
   SU(3,3) \supset Sp(6,\mathbb{R}) 
   \eea
 The quasiconformal realizations of the minimal unitary representations of the Lie algebras of these groups have the following 5-graded decompositions
 \bea
 \mathfrak{so}^*(12) = 1 \oplus (8,2) \oplus  [ \mathfrak{so}(6,2)+ \mathfrak{su}(2)+ \mathfrak{so}(1,1) ]\oplus (8,2) \oplus 1 \\
  \mathfrak{su}(3,3) = 1 \oplus (\bar{4}) \oplus [ \mathfrak{su}(2,2)+ \mathfrak{u}(1)+ \mathfrak{so}(1,1) ]\oplus (4 ) \oplus 1 \\
   \mathfrak{sp}(6,\mathbb{R}) = 1 \oplus (4) \oplus [ \mathfrak{sp}(4,\mathbb{R}) + \mathfrak{so}(1,1) ]\oplus (4 ) \oplus 1  \label{compact_0} 
 \eea
 We note that the Lie algebras of the conformal groups in the lower critical dimensions $d=6,4,3$ that appear in the grade zero subalgebra satisfy  the special isomorphisms:
 \bea
 \mathfrak{sp}(4,\mathbb{R})  \approx \mathfrak{so} (3,2) \\
  \mathfrak{su}(2,2)  \approx \mathfrak{so} (4,2) \\
   \mathfrak{so}^*(8)  \approx \mathfrak{so} (6,2) 
   \eea
 Their compact three gradings are as follows:
 \bea
 \mathfrak{so}^*(12) &=& \bar{15} \oplus \mathfrak{u}(6) \oplus 15 \\
  \mathfrak{su}(3,3) &=& (\bar{3} , \bar{3} ) \oplus [\mathfrak{su}(3)+\mathfrak{su}(3) + \mathfrak{u} (1) ]\oplus (3,3)\\
  \mathfrak{sp}(6,\mathbb{R}) &= & \bar{6} \oplus \mathfrak{u} (3) \oplus 6 
  \eea 
 The raising operators in compact three grading above decompose as follows:
 \bea
 15 &=& (6,1) + (4,2) + (1,1) \quad \Leftrightarrow  \quad SU(6) \supset SU(4)\times SU(2) \\ \nonumber
 (3,3) &=& (2,2) + (2,1) + (1,2) + (1,1)  \quad \Leftrightarrow  \quad SU(3)\times SU(3) \supset SU(2)\times SU(2) \\ \nonumber
 6 &=& 3 +2 +1 \quad    \Leftrightarrow  \quad SU(3) \supset SU(2) 
 \eea
Bosonic supersymmetry generators correspond to $(4,2) $ of $SU(4)\times SU(2)$ for the  quaternionic case, to $(2,1)+(1,2)$ of $SU(2)\times SU(2)$ in the complex case and to $2$ of $SU(2)$ in the real case.


\providecommand{\href}[2]{#2}\begingroup\raggedright\endgroup

 \end{document}